 \documentclass[final,5p,times,twocolumn,12pt]{elsarticle}

\usepackage{graphicx}

\usepackage{epstopdf}

\usepackage{amssymb}
\usepackage{amsthm}

\usepackage{amsmath}
\usepackage{amsfonts}
\usepackage{bbm}
\usepackage[tight]{subfigure}
\usepackage{tabularx}
\biboptions{numbers,square,comma,sort&compress}

\newcounter{bla}

\newcommand{\tabincell}[2]{\begin{tabular}{@{}#1@{}}#2\end{tabular}}
\input{Qcircuit}
\usepackage{wasysym}
\newcommand{\controlboth}{*-<0.27em>{\rotatebox[origin=c]{-45}{\scriptsize{$\RIGHTcircle$}}}}
\newcommand{\ctrlboth}[1]{\controlboth \qwx[#1] \qw}

\usepackage{setspace}
\usepackage{longtable}

\journal{ }

\begin{document}

\begin{frontmatter}

\title{\emph{Qcompiler}: quantum compilation with CSD method}

\author[uwa]{Y. G. Chen}

\author[uwa]{J. B. Wang\corref{author}}
\ead{jingbo.wang@uwa.edu.au}

\cortext[author] {Corresponding author}
\address[uwa]{School of Physics, The University of Western Australia, Crawley WA 6009}

\begin{abstract}
In this paper, we present a general quantum computation compiler, which maps any given quantum algorithm to a quantum circuit consisting a sequential set of elementary quantum logic gates based on recursive cosine-sine decomposition.  The resulting quantum circuit diagram is provided by directly linking the package output written in LaTex to Qcircuit.tex $<$http://www.cquic.org/Qcircuit$>$.  We illustrate the use of the \emph{Qcompiler} package through various examples with full details of the derived quantum circuits.  Besides its generality and simplicity, \emph{Qcompiler} produces quantum circuits which reflect the symmetry of the systems under study.  
\vspace{2cm}

\end{abstract}


\end{frontmatter}

\onecolumn



\noindent
{\bf Program summary} \\
\begin{small}
{\em Program Title:} \emph{Qcompiler}                                          \\
{\em Journal Reference:}                                      \\
{\em Catalogue identifier:}                                   \\
{\em Licensing provisions:} none                                  \\
{\em Programming language:} Fortran                                  \\
{\em Computer:}  any computer with a Fortran compiler                                             \\
{\em Operating system:}  Linux, Mac OS X 10.5 (and later)                                     \\
{\em RAM:}  depend on the size of the unitary matrix to be decomposed                                           
{\em Classification:}                                          \\
{\em External routines/libraries:}  Lapack                     \\
{\em Nature of problem:}\\
Decompose any given unitary operation into a quantum circuit with only elementary quantum logic gates.
   \\
{\em Solution method:}\\
This package decomposes an arbitrary unitary matrix, by applying the CSD algorithm recursively, into a series of block-diagonal matrices, which can then be readily associated with elementary quantum gates to form a quantum circuit.  \\
{\em Restrictions:}\\
The only limitation is imposed by the available memory on the user's computer.   \\
{\em Comments:}\\
This package is applicable for any \emph{arbitrary} unitary matrices, both real and complex.  If the unitary matrix is real, its corresponding quantum circuit is much simpler with only half number of quantum gates in comparison with complex matrices of the same size.   \\
{\em Running time:}\\
Memory and CPU time requirements depend critically on the size of the unitary matrix to be decomposed.  Most examples presented in this paper require a few minutes of CPU time on Intel Pentium Dual Core 2 Duo E2200 @ 2.2GHz.   \\
\end{small}


\section{\label{sec:intro}Introduction}

Quantum computing exploits the nature of the quantum world in a way that promises to solve 
problems which are intractable using conventional computers \cite{Deutsch1989, Shor1997, Nielsen2000}.  At the heart of a quantum computer lies a set of qubits whose states are manipulated by a series of elementary quantum logic gates, namely a quantum circuit, to provide the ultimate computational results.  In our earlier work \cite{Loke2011}, we developed a highly efficient quantum computation simulator to assist on the analysis of complicated quantum circuits comprised of qubit and qudit quantum gates.  In this paper, we present a quantum computation compiler which maps any given quantum algorithm to a quantum circuit consisting a set of elementary quantum logic gates. 

In the seminal papers by Barenco \emph{et al} and Deutsch \emph{et al} \cite{Barenco1995, Deutsch1995}, it was proven that any arbitrarily complex unitary operation can be implemented by a quantum circuit involving only one- or two-qubit elementary quantum logic gates.  Earlier studies applied the standard triangularization or QR-factorization scheme with Givens rotations and Gray codes to map a quantum algorithm to a series of elementary gate operations\cite{Barenco1995, Deutsch1995, Nielsen2000, Cybenko2001}.  Subsequently, a more efficient and versatile scheme based on the cosine-sine decomposition was proposed and utilized \cite{Tucci1999, Mottonen2004, Bergholm2005, Shende2005, Khan2006, Manouchehri2009}.  More recently, De Vos  \emph{et al}  \cite{DeVos2009, DeVos2011} examined another decomposition scheme, namely the Birkhoff decomposition, which was found to provide simpler quantum circuits for certain types of unitary matrices than the cosine-sine decomposition.  However, the Birkhoff decomposition does not work for general unitary matrices.  

In this work we develop a general quantum compiler, named the \emph{Qcompiler}, based on the cosine-sine decomposition scheme, because it works for arbitrary unitary matrices, it is extremely adaptable, and the core CSD code is now available as part of the LAPACK package.  Furthermore, we have managed to half the number of quantum gates in the circuit, if the quantum algorithm involves only real unitary matrices. For the \emph{Qcompiler}, the input file contains a user specified unitary matrix $U$.  \emph{Qcompiler} applies the cosine-sine decomposition (CSD) recursively and compiles $U$ into a quantum circuit with a sequence of controlled/uncontrolled rotation and phase gates. Its output contains the complete information on these elementary quantum gates, including a separate LaTex document which can be directly linked to Qcircuit.tex $<$http://www.cquic.org/Qcircuit$>$ to produce the resulting quantum circuit diagram.

This paper is structured as follows. Section 2 describes the recursive cosine-sine decomposition scheme, which maps an arbitrary unitary matrix $U$ into a quantum circuit with only one- and two-qubit logic gates. In Section 3, we provide a more cost-efficient method for real unitary matrices, which significantly reduces the number of quantum gates in the final circuit.  In Section 4, we discuss the general structure as well as the usage of the \emph{Qcompiler} package. In Section 5, we present various examples with full details and discussions of the resulting quantum circuits.  Section 6 concludes the paper.


\section{\label{sec:csd}The recursive CSD scheme: General Unitary Matrices}
The Cosine Sine Decomposition (CSD) algorithm as described in \cite{Paige1994, Sutton2009} decomposes an arbitrary ${{2}^{n}}\times {{2}^{n}}$ unitary matrix $U$ as the following
\begin{equation}
U=
\left( \begin{matrix}
   {u} & {}  \\
   {} & {v}  \\
\end{matrix} \right)
\left( \begin{matrix}
   C & S  \\
   -S & C  \\
\end{matrix} \right)
\left( \begin{matrix}
   {x} & {}  \\
   {} & {y}  \\
\end{matrix} \right),
\label{eq:csddef1}
\end{equation}
where 
\[
\begin{split}
&C=\underset{l=1,...,{{2}^{n-1}}}{\mathop{diag}}\,(\cos {{\theta }_{l}})
=\left( \begin{matrix}
   \cos {{\theta }_{1}} & {} & {} & {}  \\
   {} & \cos {{\theta }_{2}} & {} & {}  \\
   {} & {} & \ddots  & {}  \\
   {} & {} & {} & \cos {{\theta }_{{{2}^{n-1}}}}  \\
\end{matrix} \right)  
\end{split}
\]
and
\[
\begin{split}
&S=\underset{l=1,...,{{2}^{n-1}}}{\mathop{diag}}\,(\sin {{\theta }_{l}})
=\left( \begin{matrix}
   \sin {{\theta }_{1}} & {} & {} & {}  \\
   {} & \sin {{\theta }_{2}} & {} & {}  \\
   {} & {} & \ddots  & {}  \\
   {} & {} & {} & \sin {{\theta }_{{{2}^{n-1}}}}  \\
\end{matrix} \right)
\end{split}
\]
are ${{2}^{n-1}}\times {{2}^{n-1}}$ diagonal matrices, while 
$u, v, x$ and $y$ are ${{2}^{n-1}}\times {{2}^{n-1}}$ unitary matrices which can be further decomposed by the CSD algorithm recursively into a string of block diagonal unitary matrices \cite{Mottonen2004}.

To show the general structure of the decomposition, we denote all sub-matrices as $u^i_{dim}$ ignoring their explicitly different numerical values, where $i$ indicates the level of recursion and $dim$ the matrix dimension. 

At the first level, we have
\begin{equation}
\begin{split}
U=&
\left( \begin{matrix}
   u_{{{2}^{n-1}}\times {{2}^{n-1}}}^{1} & {}  \\
   {} & u_{{{2}^{n-1}}\times {{2}^{n-1}}}^{1}  \\
\end{matrix} \right)
\left( \begin{matrix}
   C_{{{2}^{n-1}}\times {{2}^{n-1}}}^{1} & S_{{{2}^{n-1}}\times {{2}^{n-1}}}^{1}  \\
   -S_{{{2}^{n-1}}\times {{2}^{n-1}}}^{1} & C_{{{2}^{n-1}}\times {{2}^{n-1}}}^{1}  \\
\end{matrix} \right)   \\
&
\left( \begin{matrix}
   u_{{{2}^{n-1}}\times {{2}^{n-1}}}^{1} & {}  \\
   {} & u_{{{2}^{n-1}}\times {{2}^{n-1}}}^{1}  \\
\end{matrix} \right)  \\
=&U^{1}A^{1}U^{1} .
\end{split}
\label{eq:csddef2}
\end{equation}

At the second level,
\[
\begin{split}
u_{{{2}^{n-1}}\times {{2}^{n-1}}}^{1}=&
\left( \begin{matrix}
   u_{{{2}^{n-2}}\times {{2}^{n-2}}}^{2} & {}  \\
   {} & u_{{{2}^{n-2}}\times {{2}^{n-2}}}^{2}  \\
\end{matrix} \right)
\left( \begin{matrix}
   C_{{{2}^{n-2}}\times {{2}^{n-2}}}^{2} & S_{{{2}^{n-2}}\times {{2}^{n-2}}}^{2}  \\
   -S_{{{2}^{n-2}}\times {{2}^{n-2}}}^{2} & C_{{{2}^{n-2}}\times {{2}^{n-2}}}^{2}  \\
\end{matrix} \right)   \\
&
\left( \begin{matrix}
   u_{{{2}^{n-2}}\times {{2}^{n-2}}}^{2} & {}  \\
   {} & u_{{{2}^{n-2}}\times {{2}^{n-2}}}^{2}  \\
\end{matrix} \right),
\end{split}
\]

\begin{equation}
\begin{split}
U^{1}=&
\left( \begin{matrix}
   u_{{{2}^{n-1}}\times {{2}^{n-1}}}^{1} & {}  \\
   {} & u_{{{2}^{n-1}}\times {{2}^{n-1}}}^{1}  \\
\end{matrix} \right) \\ 
=&
\left( \begin{matrix}
   u_{{{2}^{n-2}}\times {{2}^{n-2}}}^{2} & {} & {} & {}  \\
   {} & u_{{{2}^{n-2}}\times {{2}^{n-2}}}^{2} & {} & {}  \\
   {} & {} & u_{{{2}^{n-2}}\times {{2}^{n-2}}}^{2} & {}  \\
   {} & {} & {} & u_{{{2}^{n-2}}\times {{2}^{n-2}}}^{2}  \\
\end{matrix} \right) \\
&
\left( \begin{matrix}
   C_{{{2}^{n-2}}\times {{2}^{n-2}}}^{2} & S_{{{2}^{n-2}}\times {{2}^{n-2}}}^{2} & {} & {}  \\
   -S_{{{2}^{n-2}}\times {{2}^{n-2}}}^{2} & C_{{{2}^{n-2}}\times {{2}^{n-2}}}^{2} & {} & {}  \\
   {} & {} & C_{{{2}^{n-2}}\times {{2}^{n-2}}}^{2} & S_{{{2}^{n-2}}\times {{2}^{n-2}}}^{2}  \\
   {} & {} & -S_{{{2}^{n-2}}\times {{2}^{n-2}}}^{2} & C_{{{2}^{n-2}}\times {{2}^{n-2}}}^{2}  \\
\end{matrix} \right) \\
&
\left( \begin{matrix}
   u_{{{2}^{n-2}}\times {{2}^{n-2}}}^{2} & {} & {} & {}  \\
   {} & u_{{{2}^{n-2}}\times {{2}^{n-2}}}^{2} & {} & {}  \\
   {} & {} & u_{{{2}^{n-2}}\times {{2}^{n-2}}}^{2} & {}  \\
   {} & {} & {} & u_{{{2}^{n-2}}\times {{2}^{n-2}}}^{2}  \\
\end{matrix} \right) \\
=&U^{2}A^{2}U^{2} ,
\end{split}
\end{equation}

and 

\begin{equation}
U= U^{1}A^{1}U^{1} 
= U^{2}A^{2}U^{2}A^{1}U^{2}A^{2}U^{2} .
\end{equation}

At the $i$th level of recursion, the matrix $U^{i-1}$ is decomposed as the following:
\begin{equation}
U^{i-1}=U^{i}A^{i}U^{i}
\end{equation}
where
\begin{equation}
U^{i}=\left( \begin{matrix}
   u_{{{2}^{n-i}}\times {{2}^{n-i}}}^{i} & {} & {} & {}  \\
   {} & u_{{{2}^{n-i}}\times {{2}^{n-i}}}^{i} & {} & {}  \\
   {} & {} & \ddots & {}  \\
   {} & {} & {} & u_{{{2}^{n-i}}\times {{2}^{n-i}}}^{i}  \\
\end{matrix} \right) \\
\end{equation}

and
\begin{equation}
A^{i}=\left( \begin{matrix}
   C_{{{2}^{n-i}}\times {{2}^{n-i}}}^{i} & S_{{{2}^{n-i}}\times {{2}^{n-i}}}^{i} & {} & {} & {}  \\
   -S_{{{2}^{n-i}}\times {{2}^{n-i}}}^{i} & C_{{{2}^{n-i}}\times {{2}^{n-i}}}^{i} & {} & {} & {}  \\
   {} & {} & \ddots & {} & {}  \\ 
   {} & {} & {} & C_{{{2}^{n-i}}\times {{2}^{n-i}}}^{i} & S_{{{2}^{n-i}}\times {{2}^{n-i}}}^{i}  \\
   {} & {} & {} & -S_{{{2}^{n-i}}\times {{2}^{n-i}}}^{i} & C_{{{2}^{n-i}}\times {{2}^{n-i}}}^{i}  \\
\end{matrix} \right) .
\label{eq:Ai}
\end{equation}

At the end of the recursive process, i.e. the $n$th level, we obtain 
\begin{equation}
\begin{split}
U=&U^{1}A^{1}U^{1} \\
=&U^{2}A^{2}U^{2}A^{1}U^{2}A^{2}U^{2} \\
=&U^{3}A^{3}U^{3}A^{2}U^{3}A^{3}U^{3}A^{1}U^{3}A^{3}U^{3}A^{2}U^{3}A^{3}U^{3} \\ 
=&\cdots \cdots  \\ 
=&\left( \prod\limits_{p=1}^{{{2}^{n}}-1}{U_{p}^{n}A_{p}^{i(p)}} \right)U_{{{2}^{n}}}^{n}  ,
\end{split}
\label{eq:finaldecomposition}
\end{equation}
where $p$ marks the position of the matrix sequence, $i$ is implicitly determined by $p$, $A$ is given by Eq. (\ref{eq:Ai}), and 
\begin{equation}
U_{p}^{n}=\underset{k=1,...,{{2}^{n}}}{\mathop{diag}}\,\,(u_{p,k}^{n})=\underset{k=1,...,{{2}^{n}}}{\mathop{diag}}\,\,(\exp (i{{\varphi }_{p,k}})) .
\label{eq:DiagonalU}
\end{equation}
The above described recursive CSD scheme works for $2^n \times 2^n$ unitary matrices.  For an arbitrary $N \times N$ unitary matrix $U$, where ${{2}^{n-1}}<N\le {{2}^{n}}$, we add an identity matrix to $U$ to form a new unitary matrix 
\begin{equation}
W_{2^n \times 2^n}=
\left( \begin{matrix}
   U_{N \times N} & {}  \\
   {} & I_{(2^n-N) \times (2^n-N)}  \\
\end{matrix} \right),
\end{equation}
and then apply the recursive CSD decomposition to $W$ as described above. 

The decomposed matrices given by Eq. (\ref{eq:finaldecomposition}) can be directly related to elementary quantum gates, in particular, the phase gate $\Phi$ and the controlled rotation gate \cite{Nielsen2000, Mottonen2004}, where the rotation operation is defined as 
\begin{equation}
{{R}_{\mathbf{a}}}(\rho )=\exp (i\mathbf{a}\cdot \mathbf{\sigma }\frac{\rho }{2})=I\cos \frac{\rho }{2}+i\mathbf{a}\cdot \mathbf{\sigma }\sin \frac{\rho }{2} .
\end{equation}
If the rotation axis is $y$ or $z$, we have 
\begin{equation}
\begin{split}
{{R}_{y}}(\rho )={{R}_{y}}(2\theta )=&\exp (i{{\sigma }_{y}}\theta )=I\cos \theta +i{{\sigma }_{y}}\sin \theta \\
=&\left( \begin{matrix}
   \cos \theta  & \sin \theta   \\
   -\sin \theta  & \cos \theta   \\
\end{matrix} \right)  \\ 
\end{split}
\end{equation}
and 
\begin{equation}
\begin{split}
{{R}_{z}}(\rho )={{R}_{z}}(2\phi )=&\exp (i{{\sigma }_{z}}\phi )=I\cos \phi +i{{\sigma }_{z}}\sin \phi \\
=&\left( \begin{matrix}
   \exp (i\phi ) & 0  \\
   0 & \exp (-i\phi )  \\
\end{matrix} \right) , \\ 
\end{split}
\end{equation}
respectively.

To establish a mapping to elementary quantum gates, M\"ott\"onen \emph{et al} \cite{Mottonen2004} inserted identity matrices $I=P_{p}^{i(p)}{{(P_{p}^{i(p)})}^{\dagger }}$  after each $A_{p}^{i(p)}$ in Eq. (\ref{eq:finaldecomposition}) with which $P_{p}^{i(p)}$ commutes, and thus
\begin{equation}
U_{p}^{n}A_{p}^{i(p)}=U_{p}^{n}A_{p}^{i(p)}P_{p}^{i(p)}{{(P_{p}^{i(p)})}^{\dagger }}=U_{p}^{n}P_{p}^{i(p)}A_{p}^{i(p)}{{(P_{p}^{i(p)})}^{\dagger }}.
\label{eq:insertP}
\end{equation}
The final decomposition becomes 
\begin{equation}
U=\left( \prod\limits_{p=1}^{{{2}^{n}}-1}{B_{p}^{i(p)}A_{p}^{i(p)}} \right)\tilde{U}_{{{2}^{n}}}^{n},
\label{eq:finalCSD}
\end{equation}
where $\tilde{U}_{p}^{n}={{(P_{p-1}^{i(p-1)})}^{\dagger }} U_{p}^{n}$, except $\tilde{U}_{1}^{n}=U_1^n$, and
$B_{p}^{i(p)}=\tilde{U}_{p}^{n}P_{p}^{i(p)}$ is a $2^n\times 2^n$ diagonal unitary matrix. With a set of $P_{p}^{i(p)}$ specified by solving a set of linear equations, $B_{p}^{i(p)}$ will have the required symmetry to be equivalent to gate
\begin{equation}
{{C}^{n-1}}{{R}_{z}}(i; 1,...,i-1,i+1,...,n),
\end{equation}
where $i$ denotes the target qubit with $n-1$ control qubits.
Similarly, $A_{p}^{i}$ can be mapped to gate
\begin{equation}
{{C}^{n-1}}{{R}_{y}}(i; 1,...,i-1,i+1,...,n),
\end{equation}
and $\tilde{U}_{{2}^{n}}^{n}$ is equivalent to a series of gates
\begin{equation}
\begin{split}
&\Phi(1; )\otimes {{I}_{{{2}^{n-1}}\times {{2}^{n-1}}}}  ,\\
&{{R}_{z}}(1; )\otimes {{I}_{{{2}^{n-1}}\times {{2}^{n-1}}}} , \\
&{{C}^{1}}{{R}_{z}}(2; 1)\otimes {{I}_{{{2}^{n-2}}\times {{2}^{n-2}}}} , \\
&... \\
&{{C}^{n-2}}{{R}_{z}}(n-1; 1,...,n-2)\otimes {{I}_{2\times 2}} ,\\
&{{C}^{n-1}}{{R}_{z}}(n; 1,...,n-1) .\\
\end{split}
\end{equation}

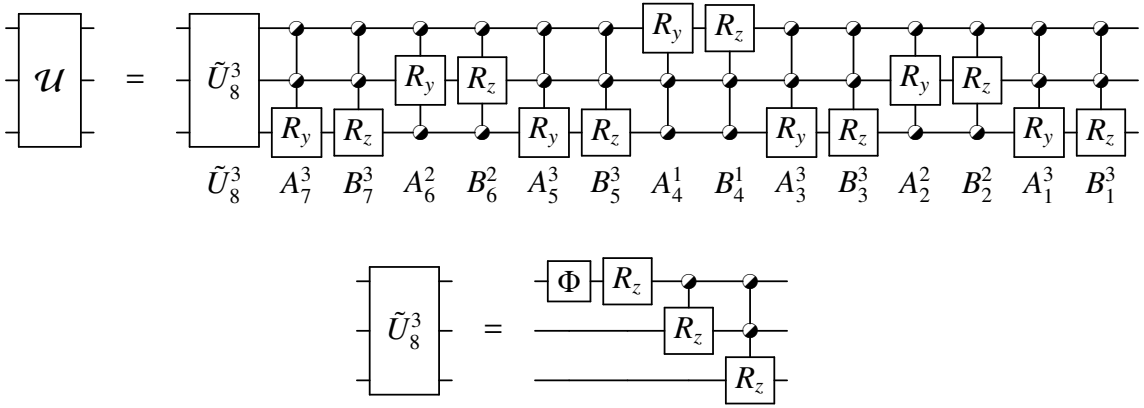
\begin{figure*}[!h]
\centering
\[
\Qcircuit @C=0.4em @R=0.1em @!R{
& \multigate{2}{\mathcal{U}} & \qw &                                          & & \multigate{2}{\mathcal{\text{$\tilde{U}_{8}^{3}$}}} & \ctrlboth{1} & \ctrlboth{1} & \ctrlboth{1} & \ctrlboth{1} & \ctrlboth{1} & \ctrlboth{1} & \gate{R_y}   & \gate{R_z}   & \ctrlboth{1} & \ctrlboth{1} & \ctrlboth{1} & \ctrlboth{1} & \ctrlboth{1} & \ctrlboth{1} & \qw \\
& \ghost{\mathcal{U}}        & \qw & \push{\rule{.3em}{0em}=\rule{.3em}{0em}} & & \ghost{\mathcal{\text{$\tilde{U}_{8}^{3}$}}}        & \ctrlboth{1} & \ctrlboth{1} & \gate{R_y}   & \gate{R_z}   & \ctrlboth{1} & \ctrlboth{1} & \ctrlboth{-1}& \ctrlboth{-1}& \ctrlboth{1} & \ctrlboth{1} & \gate{R_y}   & \gate{R_z}   & \ctrlboth{1} & \ctrlboth{1} & \qw \\
& \ghost{\mathcal{U}}        & \qw &                                          & & \ghost{\mathcal{\text{$\tilde{U}_{8}^{3}$}}}        & \gate{R_y}   & \gate{R_z}   & \ctrlboth{-1}& \ctrlboth{-1}& \gate{R_y}   & \gate{R_z}   & \ctrlboth{-1}& \ctrlboth{-1}& \gate{R_y}   & \gate{R_z}   & \ctrlboth{-1}& \ctrlboth{-1}& \gate{R_y}   & \gate{R_z}   & \qw \\
&                            &     &                                          & & \tilde{U}_{8}^{3}                                   & A_{7}^{3}    & B_{7}^{3}    & A_{6}^{2}    & B_{6}^{2}    & A_{5}^{3}    & B_{5}^{3}    & A_{4}^{1}    & B_{4}^{1}    & A_{3}^{3}    & B_{3}^{3}    & A_{2}^{2}    & B_{2}^{2}    & A_{1}^{3}    & B_{1}^{3}    &     \\
}
\]

\[
\Qcircuit @C=0.4em @R=0.1em @!R{
& \multigate{2}{\mathcal{\text{$\tilde{U}_{8}^{3}$}}} & \qw &                                          & & \gate{\Phi} & \gate{R_z} & \ctrlboth{1} & \ctrlboth{1} & \qw \\
& \ghost{\mathcal{\text{$\tilde{U}_{8}^{3}$}}}        & \qw & \push{\rule{.3em}{0em}=\rule{.3em}{0em}} & & \qw         & \qw        & \gate{R_z}   & \ctrlboth{1} & \qw \\
& \ghost{\mathcal{\text{$\tilde{U}_{8}^{3}$}}}        & \qw &                                          & & \qw         & \qw        & \qw          & \gate{R_z}   & \qw \\
}
\]
\caption{\label{fig:complex8by8}Quantum circuit for an 8-by-8 complex unitary matrix.}
\end{figure*}

As an example, we decompose a ${{2}^{3}} \times {{2}^{3}}$ unitary matrix by applying the CSD scheme recursively, i.e.   
\begin{equation}
U=B_{1}^{3}A_{1}^{3}B_{2}^{2}A_{2}^{2}B_{3}^{3}A_{3}^{3}B_{4}^{1}A_{4}^{1}B_{5}^{3}A_{5}^{3}B_{6}^{2}A_{6}^{2}B_{7}^{3}A_{7}^{3}\tilde{U}_{8}^{3}, 
\label{eq:firstU}
\end{equation}
where 
\begin{equation}
\begin{split}
A_{p}^{1}(p=4) & \equiv   {{C}^{2}}{{R}_{y}}(1; 2,3) , \\
A_{p}^{2}(p=2,6) & \equiv   {{C}^{2}}{{R}_{y}}(2; 1,3) ,\\
A_{p}^{3}(p=1,3,5,7) & \equiv   {{C}^{2}}{{R}_{y}}(3; 1,2), \\
B_{p}^{1}(p=4) & \equiv   {{C}^{2}}{{R}_{z}}(1; 2,3) ,\\
B_{p}^{2}(p=2,6) & \equiv   {{C}^{2}}{{R}_{z}}(2; 1,3), \\
B_{p}^{3}(p=1,3,5,7) & \equiv   {{C}^{2}}{{R}_{z}}(3; 1,2), \\
\end{split}
\end{equation}
and $U_{8}^{3}  \equiv  \Phi(1; ), {{R}_{z}}(1; ), {{C}^{1}}{{R}_{z}}(2; 1), {{C}^{2}}{{R}_{z}}(3; 1,2)$.  For simplicity, the identity matrices are omitted in the above expressions.  The equivalent quantum circuit for $U$ given by Eq.~\ref{eq:firstU} is shown in Fig.~\ref{fig:complex8by8}. \\

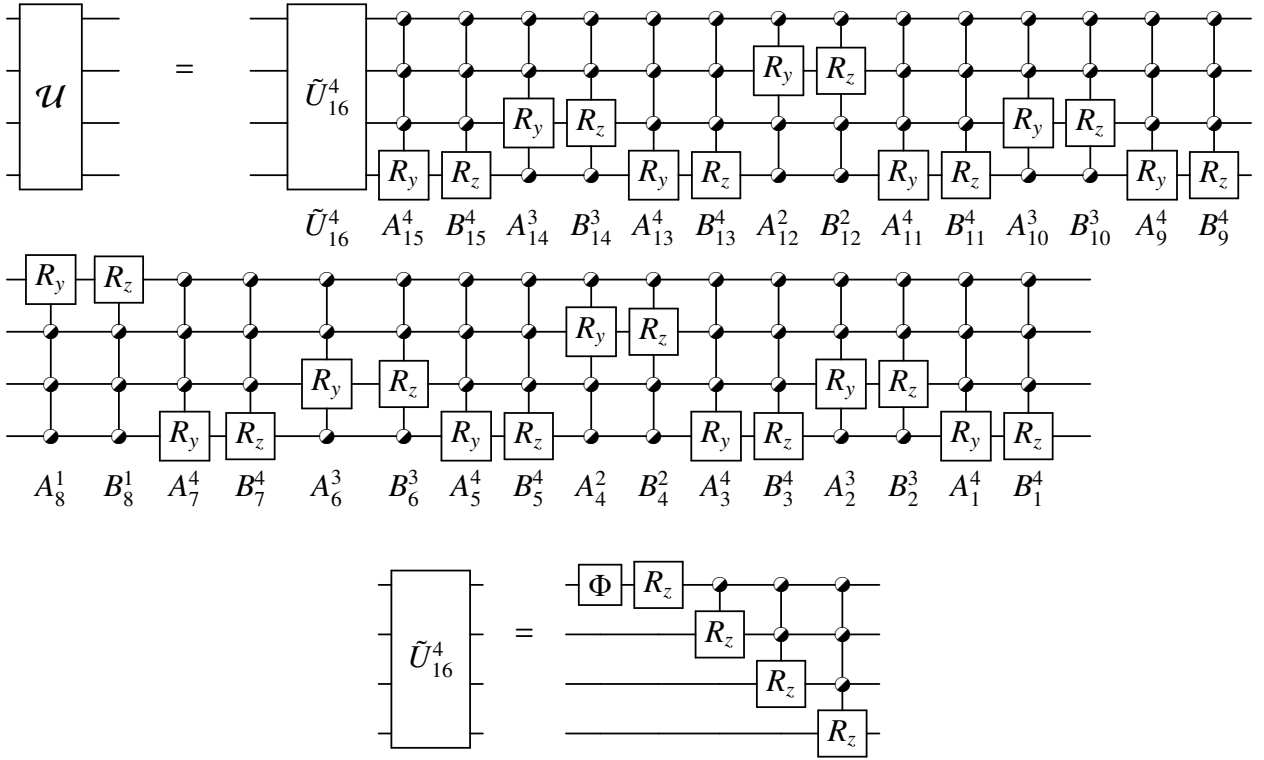
\begin{figure*}[!h]
\centering
\[
\Qcircuit @C=0.4em @R=0.1em @!R{
& \multigate{3}{\mathcal{U}} & \qw &                                          & & \multigate{3}{\mathcal{\text{$\tilde{U}_{16}^{4}$}}} & \ctrlboth{1} & \ctrlboth{1} & \ctrlboth{1} & \ctrlboth{1} & \ctrlboth{1} & \ctrlboth{1} & \ctrlboth{1} & \ctrlboth{1}  & \ctrlboth{1} & \ctrlboth{1} & \ctrlboth{1} & \ctrlboth{1} & \ctrlboth{1} & \ctrlboth{1} & \qw \\
& \ghost{\mathcal{U}}        & \qw & \push{\rule{.3em}{0em}=\rule{.3em}{0em}} & & \ghost{\mathcal{\text{$\tilde{U}_{16}^{4}$}}}        & \ctrlboth{1} & \ctrlboth{1} & \ctrlboth{1} & \ctrlboth{1} & \ctrlboth{1} & \ctrlboth{1} & \gate{R_y}   & \gate{R_z}    & \ctrlboth{1} & \ctrlboth{1} & \ctrlboth{1} & \ctrlboth{1} & \ctrlboth{1} & \ctrlboth{1} & \qw \\
& \ghost{\mathcal{U}}        & \qw &                                          & & \ghost{\mathcal{\text{$\tilde{U}_{16}^{4}$}}}        & \ctrlboth{1} & \ctrlboth{1} & \gate{R_y}   & \gate{R_z}   & \ctrlboth{1} & \ctrlboth{1} & \ctrlboth{-1}& \ctrlboth{-1} & \ctrlboth{1} & \ctrlboth{1} & \gate{R_y}   & \gate{R_z}   & \ctrlboth{1} & \ctrlboth{1} & \qw \\
& \ghost{\mathcal{U}}        & \qw &                                          & & \ghost{\mathcal{\text{$\tilde{U}_{16}^{4}$}}}        & \gate{R_y}   & \gate{R_z}   & \ctrlboth{-1}& \ctrlboth{-1}& \gate{R_y}   & \gate{R_z}   & \ctrlboth{-1}& \ctrlboth{-1} & \gate{R_y}   & \gate{R_z}   & \ctrlboth{-1}& \ctrlboth{-1}& \gate{R_y}   & \gate{R_z}   & \qw \\
&                            &     &                                          & & \tilde{U}_{16}^{4}                                   & A_{15}^{4}   & B_{15}^{4}   & A_{14}^{3}   & B_{14}^{3}   & A_{13}^{4}   & B_{13}^{4}   & A_{12}^{2}   & B_{12}^{2}    & A_{11}^{4}   & B_{11}^{4}   & A_{10}^{3}   & B_{10}^{3}   & A_{9}^{4}    & B_{9}^{4}    &     \\
& \gate{R_y}   & \gate{R_z}   & \ctrlboth{1} & \ctrlboth{1} & \ctrlboth{1} & \ctrlboth{1} & \ctrlboth{1} & \ctrlboth{1} & \ctrlboth{1} & \ctrlboth{1}  & \ctrlboth{1} & \ctrlboth{1} & \ctrlboth{1} & \ctrlboth{1} & \ctrlboth{1} & \ctrlboth{1} & \qw \\
& \ctrlboth{-1}& \ctrlboth{-1}& \ctrlboth{1} & \ctrlboth{1} & \ctrlboth{1} & \ctrlboth{1} & \ctrlboth{1} & \ctrlboth{1} & \gate{R_y}   & \gate{R_z}    & \ctrlboth{1} & \ctrlboth{1} & \ctrlboth{1} & \ctrlboth{1} & \ctrlboth{1} & \ctrlboth{1} & \qw \\
& \ctrlboth{-1}& \ctrlboth{-1}& \ctrlboth{1} & \ctrlboth{1} & \gate{R_y}   & \gate{R_z}   & \ctrlboth{1} & \ctrlboth{1} & \ctrlboth{-1}& \ctrlboth{-1} & \ctrlboth{1} & \ctrlboth{1} & \gate{R_y}   & \gate{R_z}   & \ctrlboth{1} & \ctrlboth{1} & \qw \\
& \ctrlboth{-1}& \ctrlboth{-1}& \gate{R_y}   & \gate{R_z}   & \ctrlboth{-1}& \ctrlboth{-1}& \gate{R_y}   & \gate{R_z}   & \ctrlboth{-1}& \ctrlboth{-1} & \gate{R_y}   & \gate{R_z}   & \ctrlboth{-1}& \ctrlboth{-1}& \gate{R_y}   & \gate{R_z}   & \qw \\
& A_{8}^{1}   & B_{8}^{1}     & A_{7}^{4}    & B_{7}^{4}    & A_{6}^{3}    & B_{6}^{3}    & A_{5}^{4}    & B_{5}^{4}    & A_{4}^{2}    & B_{4}^{2}     & A_{3}^{4}    & B_{3}^{4}    & A_{2}^{3}    & B_{2}^{3}    & A_{1}^{4}    & B_{1}^{4}    &     \\
}
\]

\[
\Qcircuit @C=0.4em @R=0.1em @!R{
& \multigate{3}{\mathcal{\text{$\tilde{U}_{16}^{4}$}}} & \qw &                                          & & \gate{\Phi} & \gate{R_z} & \ctrlboth{1} & \ctrlboth{1} & \ctrlboth{1} & \qw \\
& \ghost{\mathcal{\text{$\tilde{U}_{16}^{4}$}}}        & \qw & \push{\rule{.3em}{0em}=\rule{.3em}{0em}} & & \qw         & \qw        & \gate{R_z}   & \ctrlboth{1} & \ctrlboth{1} & \qw \\
& \ghost{\mathcal{\text{$\tilde{U}_{16}^{4}$}}}        & \qw &                                          & & \qw         & \qw        & \qw          & \gate{R_z}   & \ctrlboth{1} & \qw \\
& \ghost{\mathcal{\text{$\tilde{U}_{16}^{4}$}}}        & \qw &                                          & & \qw         & \qw        & \qw          & \qw          & \gate{R_z}   & \qw \\
}
\]
\caption{\label{fig:complex16by16}Quantum circuit for a 16-by-16 complex unitary matrix.}
\end{figure*}

For a ${{2}^{4}} \times {{2}^{4}}$ unitary matrix, we have
\begin{equation}
\begin{split}
U=&B_{1}^{4}A_{1}^{4}B_{2}^{3}A_{2}^{3}B_{3}^{4}A_{3}^{4}B_{4}^{2}A_{4}^{2}B_{5}^{4}A_{5}^{4}B_{6}^{3}A_{6}^{3}B_{7}^{4}A_{7}^{4}B_{8}^{1}A_{8}^{1}\\
&B_{9}^{4}A_{9}^{4}B_{10}^{3}A_{10}^{3}B_{11}^{4}A_{11}^{4}B_{12}^{2}A_{12}^{2}B_{13}^{4}A_{13}^{4}B_{14}^{3}A_{14}^{3}B_{15}^{4}A_{15}^{4}\tilde{U}_{16}^{4} .
\end{split}
\end{equation}
The corresponding quantum circuit is shown in Fig.~\ref{fig:complex16by16}. \\

\section{\label{sec:csdreal}The recursive CSD scheme: Real Unitary Matrices}

If the ${{2}^{n}}\times {{2}^{n}}$ unitary matrix $U$ is real, we have $u_{p,k}^{n} = \exp (i{{\varphi }_{p,k}}) = 1$ or $-1$ from Eq.~\ref{eq:DiagonalU} and $U_{p}^{n}$ becomes a diagonal matrix consisting of only 1 or $-1$.
In this case, we can insert the identity matrix $I=\tilde{U}_{p}^{n}{{(\tilde{U}_{p}^{n})}^{\dagger }}=\tilde{U}_{p}^{n}\tilde{U}_{p}^{n}$ after each $A_{p}^{i(p)}$ in Eq. ~\ref{eq:finaldecomposition}, i.e. 
\begin{equation}
U_{p}^{n}A_{p}^{i(p)}=U_{p}^{n}A_{p}^{i(p)}\tilde{U}_{p}^{n}\tilde{U}_{p}^{n}
\end{equation}
where $\tilde{U}_{p}^{n}=\prod\limits_{q=1}^{p}{U_{q}^{n}}$.
The decomposition for a real unitary $U$ then becomes
\begin{equation}
U=\left( \prod\limits_{p=1}^{{{2}^{n}}-1}{\tilde{A}_{p}^{i(p)}} \right)\tilde{U}_{{{2}^{n}}}^{n}
\end{equation}
where $\tilde{A}_{p}^{i(p)}=\tilde{U}_{p}^{n}A_{p}^{i(p)}\tilde{U}_{p}^{n}$.  Note for general unitary matrices, it requires the solution of a set of linear equations to determine each inserted $P_{p}^{i(p)}$ in Eq.~\ref{eq:insertP} \cite{Mottonen2004}.  Here, for real unitary matrices, we only need to calculate the product of ${U}_{1}^{n}, {U}_{2}^{n}, ... , {U}_{p}^{n}$ to obtain the inserted $\tilde{U}_{p}^{n}$.
Furthermore, we end up with only half number of decomposed matrices in comparison with Eq.~\ref{eq:finalCSD}.
Since $\tilde{U}_{p}^{n}$ is a diagonal matrix with only two possible values, 1 or $-1$, $\tilde{A}_{p}^{i(p)}$ can be readily mapped to the gate ${{C}^{n-1}}{{R}_{y}}(i; 1,...,i-1,i+1,...,n)$. The gates for $\tilde{A}_{p}^{i(p)}$ and ${A}_{p}^{i(p)}$ are almost the same except for the signs of some rotation angles.

The last matrix $\tilde{U}_{{2}^{n}}^{n}$ can be written as
\begin{equation}
\tilde{U}_{{2}^{n}}^{n} = \prod\limits_{m=1}^{n}{\underset{k=1,...,{{2}^{n-m}}}{\mathop{diag}}\,({{D}_{m,k}})\otimes {{I}_{{{2}^{m-1}}\times {{2}^{m-1}}}}}
\end{equation}
where ${{D}_{m,k}}$ are either 
$\left(\begin{smallmatrix}
 1 & 0 \\
 0 & 1 \\
\end{smallmatrix}\right)$
or
$\left(\begin{smallmatrix}
 1 & 0 \\
 0 & -1 \\
\end{smallmatrix}\right)$,
with the former being an identity gate and the later a $\Pi$ gate. Therefore, $U_{{2}^{n}}^{n}$ is equivalent to a \emph{subset} of the following quantum gates:
\[
\begin{split}
&\Pi(1; )\otimes {{I}_{{{2}^{n-1}}\times {{2}^{n-1}}}} \\
&{{C}^{1}}\Pi(2; 1)\otimes {{I}_{{{2}^{n-2}}\times {{2}^{n-2}}}} \\
&... \\
&{{C}^{n-2}}\Pi(n-1; 1,...,n-2)\otimes {{I}_{2\times 2}} \\
&{{C}^{n-1}}\Pi(n; 1,...,n-1) . \\
\end{split}
\]

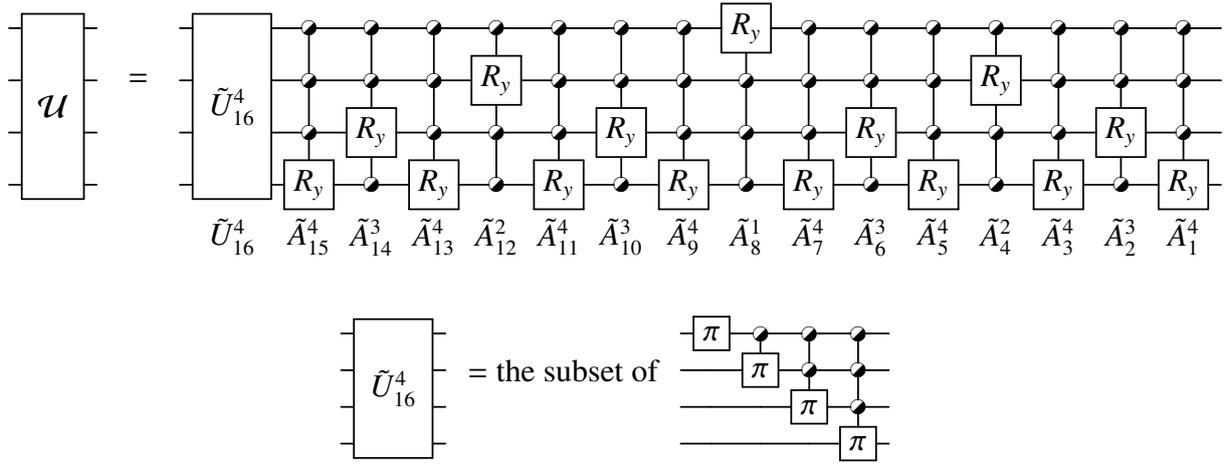
\begin{figure*}[!h]
\centering
\[
\Qcircuit @C=0.4em @R=0.1em @!R{
& \multigate{3}{\mathcal{U}} & \qw &                                          & & \multigate{3}{\mathcal{\text{$\tilde{U}_{16}^{4}$}}} & \ctrlboth{1}       & \ctrlboth{1}       & \ctrlboth{1}       & \ctrlboth{1}       & \ctrlboth{1}       & \ctrlboth{1}       & \ctrlboth{1}      & \gate{R_y}        & \ctrlboth{1}      & \ctrlboth{1}      & \ctrlboth{1}      & \ctrlboth{1}      & \ctrlboth{1}      & \ctrlboth{1}      & \ctrlboth{1}      & \qw \\
& \ghost{\mathcal{U}}        & \qw & \push{\rule{.3em}{0em}=\rule{.3em}{0em}} & & \ghost{\mathcal{\text{$\tilde{U}_{16}^{4}$}}}        & \ctrlboth{1}       & \ctrlboth{1}       & \ctrlboth{1}       & \gate{R_y}         & \ctrlboth{1}       & \ctrlboth{1}       & \ctrlboth{1}      & \ctrlboth{-1}     & \ctrlboth{1}      & \ctrlboth{1}      & \ctrlboth{1}      & \gate{R_y}        & \ctrlboth{1}      & \ctrlboth{1}      & \ctrlboth{1}      & \qw \\
& \ghost{\mathcal{U}}        & \qw &                                          & & \ghost{\mathcal{\text{$\tilde{U}_{16}^{4}$}}}        & \ctrlboth{1}       & \gate{R_y}         & \ctrlboth{1}       & \ctrlboth{-1}      & \ctrlboth{1}       & \gate{R_y}         & \ctrlboth{1}      & \ctrlboth{-1}     & \ctrlboth{1}      & \gate{R_y}        & \ctrlboth{1}      & \ctrlboth{-1}     & \ctrlboth{1}      & \gate{R_y}        & \ctrlboth{1}      & \qw \\
& \ghost{\mathcal{U}}        & \qw &                                          & & \ghost{\mathcal{\text{$\tilde{U}_{16}^{4}$}}}        & \gate{R_y}         & \ctrlboth{-1}      & \gate{R_y}         & \ctrlboth{-1}      & \gate{R_y}         & \ctrlboth{-1}      & \gate{R_y}        & \ctrlboth{-1}     & \gate{R_y}        & \ctrlboth{-1}     & \gate{R_y}        & \ctrlboth{-1}     & \gate{R_y}        & \ctrlboth{-1}     & \gate{R_y}        & \qw \\
&                            &     &                                          & & \tilde{U}_{16}^{4}                                   & \tilde{A}_{15}^{4} & \tilde{A}_{14}^{3} & \tilde{A}_{13}^{4} & \tilde{A}_{12}^{2} & \tilde{A}_{11}^{4} & \tilde{A}_{10}^{3} & \tilde{A}_{9}^{4} & \tilde{A}_{8}^{1} & \tilde{A}_{7}^{4} & \tilde{A}_{6}^{3} & \tilde{A}_{5}^{4} & \tilde{A}_{4}^{2} & \tilde{A}_{3}^{4} & \tilde{A}_{2}^{3} & \tilde{A}_{1}^{4} &     \\
}
\]

\[
\Qcircuit @C=0.4em @R=0.1em @!R{
& \multigate{3}{\mathcal{\text{$\tilde{U}_{16}^{4}$}}} & \qw &                                                               & & \gate{\pi} & \ctrlboth{1} & \ctrlboth{1} & \ctrlboth{1} & \qw \\
& \ghost{\mathcal{\text{$\tilde{U}_{16}^{4}$}}}        & \qw & \push{\rule{.3em}{0em}\text{= the subset of}\rule{.3em}{0em}} & & \qw        & \gate{\pi}   & \ctrlboth{1} & \ctrlboth{1} & \qw \\
& \ghost{\mathcal{\text{$\tilde{U}_{16}^{4}$}}}        & \qw &                                                               & & \qw        & \qw          & \gate{\pi}   & \ctrlboth{1} & \qw \\
& \ghost{\mathcal{\text{$\tilde{U}_{16}^{4}$}}}        & \qw &                                                               & & \qw        & \qw          & \qw          & \gate{\pi}   & \qw \\
}
\]
\caption{\label{fig:real16by16}The circuit of a 16-by-16 real $U$.}
\end{figure*}

As an example we decompose a 16-by-16 real unitary matrix $U$ and get
\begin{equation}
U={\tilde{A}}_{1}^{4}\tilde{A}_{2}^{3}\tilde{A}_{3}^{4}\tilde{A}_{4}^{2}\tilde{A}_{5}^{4}\tilde{A}_{6}^{3}\tilde{A}_{7}^{4}\tilde{A}_{8}^{1}\tilde{A}_{9}^{4}\tilde{A}_{10}^{3}\tilde{A}_{11}^{4}\tilde{A}_{12}^{2}\tilde{A}_{13}^{4} \tilde{A}_{14}^{3}\tilde{A}_{15}^{4}\tilde{U}_{16}^{4}
\end{equation}
Its corresponding circuit is, as shown in Fig.~\ref{fig:real16by16}, much simpler than that for a 16-by-16 complex unitary matrix.  The number of gates is reduced to half.

\section{\label{sec:software}Overview of the Software}

The \emph{Qcompiler} package, written in Fortran, consists of 1 main program and 12 subroutines. This package utilizes recursively Sutton's CSD subroutines \cite{Sutton2009}, which is now part of Lapack available at http://www.netlib.org/lapack under a permissive free software license.

\begin{figure*}[!h]
\centering
\includegraphics[scale=0.77]{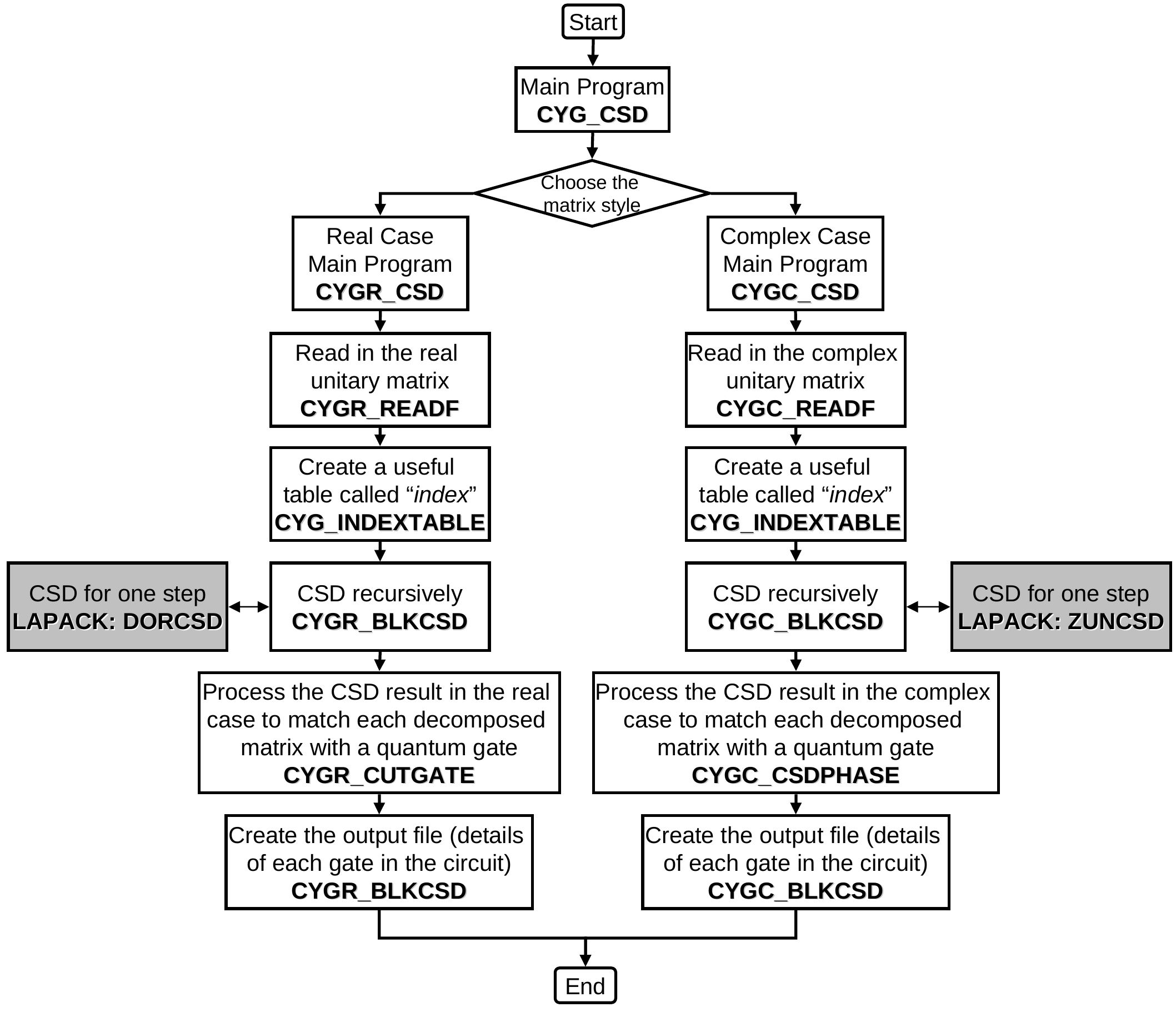}
\caption{\label{fig:flowchart}Flow chart for the \emph{Qcompiler} package.}
\end{figure*}

\begin{figure}[!h]
\centering
\includegraphics[scale=0.55]{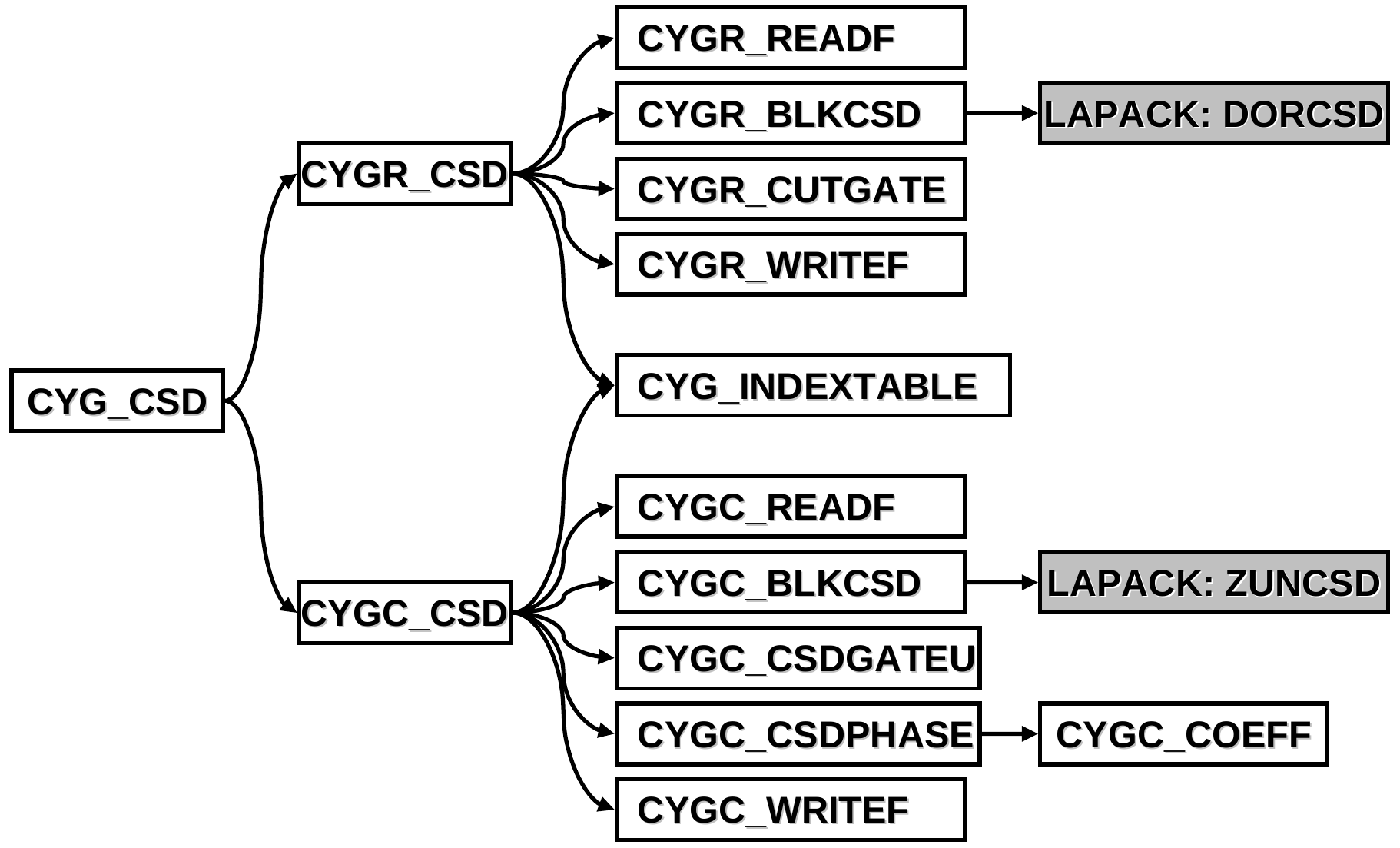}
\caption{\label{fig:callgraph}Flow chart for all subroutine calls.}
\end{figure}

As described in Section 3, the \emph{Qcompiler} package decomposes $U$ as
\begin{equation}
U=\left\{ \begin{array}{*{35}{l}}
   \left( \prod\limits_{p=1}^{{{2}^{n}}-1}{\tilde{A}_{p}^{i(p)}} \right)\tilde{U}_{{{2}^{n}}}^{n} & \text{if }U\text{ is real,}  \\
   \left( \prod\limits_{p=1}^{{{2}^{n}}-1}{B_{p}^{i(p)}A_{p}^{i(p)}} \right)\tilde{U}_{{{2}^{n}}}^{n} & \text{if }U\text{ is complex.}  \\
\end{array} \right.
\end{equation}
It then generates an output file containing a complete description of each quantum gate in the circuit. 
The principal flow chart of \emph{Qcompiler} is shown in Fig.~\ref{fig:flowchart} with all subroutines calls summarized in Fig.~\ref{fig:callgraph}.  

\newpage

\newpage

\section{\label{sec:example}Example}

\subsection{\label{sec:example complexU}Complex unitary matrix}

An 8-by-8 complex unitary matrix $U$ is generated randomly as the following: 
\begin{equation}
U=\left( \begin{smallmatrix}
   0.6501-0.3423i & -0.0792-0.1132i & -0.1411-0.0356i & -0.1727+0.1594i & -0.2945-0.2656i & -0.3970+0.0475i & -0.1703-0.0385i & -0.0757+0.1065i  \\
   -0.2339-0.1113i & 0.5429+0.0637i & -0.0551-0.0559i & -0.0454-0.1349i & -0.0212+0.0956i & -0.1494-0.0384i & -0.5056-0.1816i & -0.5333-0.0362i  \\
   -0.1885-0.1242i & -0.1468-0.2134i & 0.6018-0.1214i & -0.5436-0.1291i & -0.1374+0.0600i & -0.1848-0.1079i & 0.0026+0.1374i & 0.0116-0.3250i  \\
   -0.2882+0.1223i & -0.2745+0.1175i & -0.0532-0.2853i & 0.2171-0.4183i & -0.5067-0.4492i & -0.0320-0.0010i & -0.0223+0.0879i & -0.1129+0.1611i  \\
   0.0801-0.1293i & -0.4807+0.1042i & -0.3721+0.0264i & -0.2358-0.3746i & 0.4089+0.0434i & 0.1480+0.1193i & -0.2750+0.1932i & -0.1832-0.2196i  \\
   -0.1002+0.0211i & -0.1428-0.4283i & -0.2636+0.1747i & -0.1363+0.2767i & -0.3588-0.0596i & 0.5210-0.2426i & -0.0680-0.2148i & -0.1792-0.2124i  \\
   -0.2579-0.1785i & -0.0825-0.2606i & -0.2869+0.1512i & -0.1106-0.2008i & 0.1490+0.0484i & -0.3993-0.1107i & 0.5480-0.3288i & -0.1874+0.1716i  \\
   -0.3122+0.1212i & 0.0733+0.0072i & -0.4026+0.0719i & 0.0101+0.2156i & -0.0048-0.1621i & -0.4437-0.1859i & -0.1984+0.2056i & 0.4193-0.3917i  \\
\end{smallmatrix} \right).
\label{eq:random8by8matrix}
\end{equation}
Recursive CSD decomposition gives 
$ U=B_{1}^{3}A_{1}^{3}B_{2}^{2}A_{2}^{2}B_{3}^{3}A_{3}^{3}B_{4}^{1}A_{4}^{1}B_{5}^{3}A_{5}^{3}B_{6}^{2}A_{6}^{2}B_{7}^{3}A_{7}^{3}\tilde{U}_{8}^{3}.
$
Their corresponding quantum gates are detailed in the output file produced by \emph{Qcompiler}, as explained below.

\begin{spacing}{1}
\begin{center}
\begin{longtable}{l|l|ll}
    \hline
      Output of the package &  Matrix & \multicolumn{2}{c}{Gate} \\ 
    \hline
    \endhead
      \tabincell{l}{\\GATEPHASE\\  1;\\  0.1586} 
    & 
    & \tabincell{l}{\\$\Phi (1;)$\\$\Phi =2\pi \cdot 0.1586$} 
    & \Qcircuit @C=0.4em @R=0.1em @!R{
      & \gate{\Phi} & \qw \\
      & \qw         & \qw \\
      & \qw         & \qw \\
      \\
      }
    \\
      \tabincell{l}{\\GATEZ\\  1;\\ -0.9734} 
    & $\tilde{U}_{8}^{3}$
    & \tabincell{l}{\\${{R}_{z}}(1;)$\\$2{{{\phi }'}_{1}}=2\pi \cdot (-0.9734)$} 
    & \Qcircuit @C=0.4em @R=0.1em @!R{
      & \gate{R_z} & \qw \\
      & \qw        & \qw \\
      & \qw        & \qw \\
      \\
      }
    \\
      \tabincell{l}{\\GATEZ\\  2;  1\\ -0.0856  0.4745} 
    &
    & \tabincell{l}{\\${{C}^{1}}{{R}_{z}}(2;1)$\\$2{{{\phi }'}_{1}}=2\pi \cdot (-0.0856)$\\$2{{{\phi }'}_{2}}=2\pi \cdot 0.4745$} 
    & \Qcircuit @C=0.4em @R=0.1em @!R{
      & \ctrlboth{1} & \qw &   &   & \ctrlo{1}    & \ctrl{1}     & \qw \\
      & \gate{R_z}   & \qw & = &   & \gate{R_z^1} & \gate{R_z^2} & \qw \\
      & \qw          & \qw &   &   & \qw          & \qw          & \qw \\
      \\
      }
    \\
      \tabincell{l}{\\GATEZ\\  3;  1,  2\\  0.0975 -0.0388 -0.4445  0.0461} 
    & 
    & \tabincell{l}{\\${{C}^{2}}{{R}_{z}}(3;1,2)$\\$2{{{\phi }'}_{1}}=2\pi \cdot 0.0975$\\$2{{{\phi }'}_{2}}=2\pi \cdot (-0.0388)$\\$2{{{\phi }'}_{3}}=2\pi \cdot (-0.4445)$\\$2{{{\phi }'}_{4}}=2\pi \cdot 0.0461$} 
    & \Qcircuit @C=0.4em @R=0.1em @!R{
      & \ctrlboth{1} & \qw &   &   & \ctrlo{1}    & \ctrlo{1}    & \ctrl{1}     & \ctrl{1}     & \qw \\
      & \ctrlboth{1} & \qw & = &   & \ctrlo{1}    & \ctrl{1}     & \ctrlo{1}    & \ctrl{1}     & \qw \\
      & \gate{R_z}   & \qw &   &   & \gate{R_z^1} & \gate{R_z^2} & \gate{R_z^3} & \gate{R_z^4} & \qw \\
      \\
      }
    \\
    \hline
      \tabincell{l}{\\GATEY\\  3;  1,  2\\  0.0758  0.2318  0.4094  0.2817} 
    & $A_{7}^{3}$
    & \tabincell{l}{\\${{C}^{2}}{{R}_{y}}(3;1,2)$\\$2{{\theta }_{1}}=2\pi \cdot 0.0758$\\$2{{\theta }_{2}}=2\pi \cdot 0.2318$\\$2{{\theta }_{3}}=2\pi \cdot 0.4094$\\$2{{\theta }_{4}}=2\pi \cdot 0.2817$}
    & \Qcircuit @C=0.4em @R=0.1em @!R{
      & \ctrlboth{1} & \qw &   &   & \ctrlo{1}    & \ctrlo{1}    & \ctrl{1}     & \ctrl{1}     & \qw \\
      & \ctrlboth{1} & \qw & = &   & \ctrlo{1}    & \ctrl{1}     & \ctrlo{1}    & \ctrl{1}     & \qw \\
      & \gate{R_y}   & \qw &   &   & \gate{R_y^1} & \gate{R_y^2} & \gate{R_y^3} & \gate{R_y^4} & \qw \\
      \\
      }
    \\
    \hline
      \tabincell{l}{\\GATEZ\\  3;  1,  2\\ -0.5384  0.1700 -0.6665 -0.2091} 
    & $B_{7}^{3}$
    & \tabincell{l}{\\${{C}^{2}}{{R}_{z}}(3;1,2)$\\$2{{\phi }_{1}}=2\pi \cdot (-0.5384)$\\$2{{\phi }_{2}}=2\pi \cdot 0.1700$\\$2{{\phi }_{3}}=2\pi \cdot (-0.6665)$\\$2{{\phi }_{4}}=2\pi \cdot (-0.2091)$} 
    & \Qcircuit @C=0.4em @R=0.1em @!R{
      & \ctrlboth{1} & \qw &   &   & \ctrlo{1}    & \ctrlo{1}    & \ctrl{1}     & \ctrl{1}     & \qw \\
      & \ctrlboth{1} & \qw & = &   & \ctrlo{1}    & \ctrl{1}     & \ctrlo{1}    & \ctrl{1}     & \qw \\
      & \gate{R_z}   & \qw &   &   & \gate{R_z^1} & \gate{R_z^2} & \gate{R_z^3} & \gate{R_z^4} & \qw \\
      \\
      }
    \\
    \hline
      \tabincell{l}{\\GATEY\\  2;  1,  3\\  0.1573  0.3405  0.1019  0.4831} 
    & $A_{6}^{2}$
    & \tabincell{l}{\\${{C}^{2}}{{R}_{y}}(2;1,3)$\\$2{{\theta }_{1}}=2\pi \cdot 0.1573$\\$2{{\theta }_{2}}=2\pi \cdot 0.3405$\\$2{{\theta }_{3}}=2\pi \cdot 0.1019$\\$2{{\theta }_{4}}=2\pi \cdot 0.4831$}
    & \Qcircuit @C=0.4em @R=0.1em @!R{
      & \ctrlboth{1} & \qw &   &   & \ctrlo{1}    & \ctrlo{1}    & \ctrl{1}     & \ctrl{1}     & \qw \\
      & \gate{R_y}   & \qw & = &   & \gate{R_y^1} & \gate{R_y^2} & \gate{R_y^3} & \gate{R_y^4} & \qw \\
      & \ctrlboth{-1}& \qw &   &   & \ctrlo{-1}   & \ctrl{-1}    & \ctrlo{-1}   & \ctrl{-1}    & \qw \\
      \\
      }
    \\
    \hline
      \tabincell{l}{\\GATEZ\\  2;  1,  3\\ -0.0876 -0.2271  0.2202  0.2315} 
    & $B_{6}^{2}$
    & \tabincell{l}{\\${{C}^{2}}{{R}_{z}}(2;1,3)$\\$2{{\phi }_{1}}=2\pi \cdot (-0.0876)$\\$2{{\phi }_{2}}=2\pi \cdot (-0.2271)$\\$2{{\phi }_{3}}=2\pi \cdot 0.2202$\\$2{{\phi }_{4}}=2\pi \cdot 0.2315$} 
    & \Qcircuit @C=0.4em @R=0.1em @!R{
      & \ctrlboth{1} & \qw &   &   & \ctrlo{1}    & \ctrlo{1}    & \ctrl{1}     & \ctrl{1}     & \qw \\
      & \gate{R_z}   & \qw & = &   & \gate{R_z^1} & \gate{R_z^2} & \gate{R_z^3} & \gate{R_z^4} & \qw \\
      & \ctrlboth{-1}& \qw &   &   & \ctrlo{-1}   & \ctrl{-1}    & \ctrlo{-1}   & \ctrl{-1}    & \qw \\
      \\
      }
    \\
    \hline
      \tabincell{l}{\\GATEY\\  3;  1,  2\\  0.1231  0.3567  0.2444  0.2683} 
    & $A_{5}^{3}$
    & \tabincell{l}{\\${{C}^{2}}{{R}_{y}}(3;1,2)$\\$2{{\theta }_{1}}=2\pi \cdot 0.1231$\\$2{{\theta }_{2}}=2\pi \cdot 0.3567$\\$2{{\theta }_{3}}=2\pi \cdot 0.2444$\\$2{{\theta }_{4}}=2\pi \cdot 0.2683$}
    & \Qcircuit @C=0.4em @R=0.1em @!R{
      & \ctrlboth{1} & \qw &   &   & \ctrlo{1}    & \ctrlo{1}    & \ctrl{1}     & \ctrl{1}     & \qw \\
      & \ctrlboth{1} & \qw & = &   & \ctrlo{1}    & \ctrl{1}     & \ctrlo{1}    & \ctrl{1}     & \qw \\
      & \gate{R_y}   & \qw &   &   & \gate{R_y^1} & \gate{R_y^2} & \gate{R_y^3} & \gate{R_y^4} & \qw \\
      \\
      }
    \\
    \hline
      \tabincell{l}{\\GATEZ\\  3;  1,  2\\ -0.3146  0.4879 -0.5310 -0.1047} 
    & $B_{5}^{3}$
    & \tabincell{l}{\\${{C}^{2}}{{R}_{z}}(3;1,2)$\\$2{{\phi }_{1}}=2\pi \cdot (-0.3146)$\\$2{{\phi }_{2}}=2\pi \cdot 0.4879$\\$2{{\phi }_{3}}=2\pi \cdot (-0.5310)$\\$2{{\phi }_{4}}=2\pi \cdot (-0.1047)$} 
    & \Qcircuit @C=0.4em @R=0.1em @!R{
      & \ctrlboth{1} & \qw &   &   & \ctrlo{1}    & \ctrlo{1}    & \ctrl{1}     & \ctrl{1}     & \qw \\
      & \ctrlboth{1} & \qw & = &   & \ctrlo{1}    & \ctrl{1}     & \ctrlo{1}    & \ctrl{1}     & \qw \\
      & \gate{R_z}   & \qw &   &   & \gate{R_z^1} & \gate{R_z^2} & \gate{R_z^3} & \gate{R_z^4} & \qw \\
      \\
      }
    \\
    \hline
      \tabincell{l}{\\GATEY\\  1;  2,  3\\  0.0901  0.1279  0.2752  0.3789} 
    & $A_{4}^{1}$
    & \tabincell{l}{\\${{C}^{2}}{{R}_{y}}(1;2,3)$\\$2{{\theta }_{1}}=2\pi \cdot 0.0901$\\$2{{\theta }_{2}}=2\pi \cdot 0.1279$\\$2{{\theta }_{3}}=2\pi \cdot 0.2752$\\$2{{\theta }_{4}}=2\pi \cdot 0.3789$}
    & \Qcircuit @C=0.4em @R=0.1em @!R{
      & \gate{R_y}   & \qw &   &   & \gate{R_y^1} & \gate{R_y^2} & \gate{R_y^3} & \gate{R_y^4} & \qw \\
      & \ctrlboth{-1}& \qw & = &   & \ctrlo{-1}   & \ctrlo{-1}   & \ctrl{-1}    & \ctrl{-1}    & \qw \\
      & \ctrlboth{-1}& \qw &   &   & \ctrlo{-1}   & \ctrl{-1}    & \ctrlo{-1}   & \ctrl{-1}    & \qw \\
      \\
      }
    \\
    \hline
      \tabincell{l}{\\GATEZ\\  1;  2,  3\\  0.5219  0.5019  0.2641 -0.0838} 
    & $B_{4}^{1}$
    & \tabincell{l}{\\${{C}^{2}}{{R}_{z}}(1;2,3)$\\$2{{\phi }_{1}}=2\pi \cdot 0.5219$\\$2{{\phi }_{2}}=2\pi \cdot 0.5019$\\$2{{\phi }_{3}}=2\pi \cdot 0.2641$\\$2{{\phi }_{4}}=2\pi \cdot (-0.0838)$} 
    & \Qcircuit @C=0.4em @R=0.1em @!R{
      & \gate{R_z}   & \qw &   &   & \gate{R_z^1} & \gate{R_z^2} & \gate{R_z^3} & \gate{R_z^4} & \qw \\
      & \ctrlboth{-1}& \qw & = &   & \ctrlo{-1}   & \ctrlo{-1}   & \ctrl{-1}    & \ctrl{-1}    & \qw \\
      & \ctrlboth{-1}& \qw &   &   & \ctrlo{-1}   & \ctrl{-1}    & \ctrlo{-1}   & \ctrl{-1}    & \qw \\
      \\
      }
    \\
    \hline
      \tabincell{l}{\\GATEY\\  3;  1,  2\\  0.1316  0.2352  0.2069  0.3275} 
    & $A_{3}^{3}$
    & \tabincell{l}{\\${{C}^{2}}{{R}_{y}}(3;1,2)$\\$2{{\theta }_{1}}=2\pi \cdot 0.1316$\\$2{{\theta }_{2}}=2\pi \cdot 0.2352$\\$2{{\theta }_{3}}=2\pi \cdot 0.2069$\\$2{{\theta }_{4}}=2\pi \cdot 0.3275$}
    & \Qcircuit @C=0.4em @R=0.1em @!R{
      & \ctrlboth{1} & \qw &   &   & \ctrlo{1}    & \ctrlo{1}    & \ctrl{1}     & \ctrl{1}     & \qw \\
      & \ctrlboth{1} & \qw & = &   & \ctrlo{1}    & \ctrl{1}     & \ctrlo{1}    & \ctrl{1}     & \qw \\
      & \gate{R_y}   & \qw &   &   & \gate{R_y^1} & \gate{R_y^2} & \gate{R_y^3} & \gate{R_y^4} & \qw \\
      \\
      }
    \\
    \hline
      \tabincell{l}{\\GATEZ\\  3;  1,  2\\ -0.5367 -0.3010 -0.0056  0.5324} 
    & $B_{3}^{3}$
    & \tabincell{l}{\\${{C}^{2}}{{R}_{z}}(3;1,2)$\\$2{{\phi }_{1}}=2\pi \cdot (-0.5367)$\\$2{{\phi }_{2}}=2\pi \cdot (-0.3010)$\\$2{{\phi }_{3}}=2\pi \cdot (-0.0056)$\\$2{{\phi }_{4}}=2\pi \cdot 0.5324$} 
    & \Qcircuit @C=0.4em @R=0.1em @!R{
      & \ctrlboth{1} & \qw &   &   & \ctrlo{1}    & \ctrlo{1}    & \ctrl{1}     & \ctrl{1}     & \qw \\
      & \ctrlboth{1} & \qw & = &   & \ctrlo{1}    & \ctrl{1}     & \ctrlo{1}    & \ctrl{1}     & \qw \\
      & \gate{R_z}   & \qw &   &   & \gate{R_z^1} & \gate{R_z^2} & \gate{R_z^3} & \gate{R_z^4} & \qw \\
      \\
      }
    \\
    \hline
      \tabincell{l}{\\GATEY\\  2;  1,  3\\  0.1898  0.4103  0.2375  0.3869} 
    & $A_{2}^{2}$
    & \tabincell{l}{\\${{C}^{2}}{{R}_{y}}(2;1,3)$\\$2{{\theta }_{1}}=2\pi \cdot 0.1898$\\$2{{\theta }_{2}}=2\pi \cdot 0.4103$\\$2{{\theta }_{3}}=2\pi \cdot 0.2375$\\$2{{\theta }_{4}}=2\pi \cdot 0.3869$}
    & \Qcircuit @C=0.4em @R=0.1em @!R{
      & \ctrlboth{1} & \qw &   &   & \ctrlo{1}    & \ctrlo{1}    & \ctrl{1}     & \ctrl{1}     & \qw \\
      & \gate{R_y}   & \qw & = &   & \gate{R_y^1} & \gate{R_y^2} & \gate{R_y^3} & \gate{R_y^4} & \qw \\
      & \ctrlboth{-1}& \qw &   &   & \ctrlo{-1}   & \ctrl{-1}    & \ctrlo{-1}   & \ctrl{-1}    & \qw \\
      \\
      }
    \\
    \hline
      \tabincell{l}{\\GATEZ\\  2;  1,  3\\ -0.1607 -0.5540  0.0320  0.8280} 
    & $B_{2}^{2}$
    & \tabincell{l}{\\${{C}^{2}}{{R}_{z}}(2;1,3)$\\$2{{\phi }_{1}}=2\pi \cdot (-0.1607)$\\$2{{\phi }_{2}}=2\pi \cdot (-0.5540)$\\$2{{\phi }_{3}}=2\pi \cdot 0.0320$\\$2{{\phi }_{4}}=2\pi \cdot 0.8280$} 
    & \Qcircuit @C=0.4em @R=0.1em @!R{
      & \ctrlboth{1} & \qw &   &   & \ctrlo{1}    & \ctrlo{1}    & \ctrl{1}     & \ctrl{1}     & \qw \\
      & \gate{R_z}   & \qw & = &   & \gate{R_z^1} & \gate{R_z^2} & \gate{R_z^3} & \gate{R_z^4} & \qw \\
      & \ctrlboth{-1}& \qw &   &   & \ctrlo{-1}   & \ctrl{-1}    & \ctrlo{-1}   & \ctrl{-1}    & \qw \\
      \\
      }
    \\
    \hline
      \tabincell{l}{\\GATEY\\  3;  1,  2\\  0.1077  0.4126  0.3956  0.2628} 
    & $A_{1}^{3}$
    & \tabincell{l}{\\${{C}^{2}}{{R}_{y}}(3;1,2)$\\$2{{\theta }_{1}}=2\pi \cdot 0.1077$\\$2{{\theta }_{2}}=2\pi \cdot 0.4126$\\$2{{\theta }_{3}}=2\pi \cdot 0.3956$\\$2{{\theta }_{4}}=2\pi \cdot 0.2628$}
    & \Qcircuit @C=0.4em @R=0.1em @!R{
      & \ctrlboth{1} & \qw &   &   & \ctrlo{1}    & \ctrlo{1}    & \ctrl{1}     & \ctrl{1}     & \qw \\
      & \ctrlboth{1} & \qw & = &   & \ctrlo{1}    & \ctrl{1}     & \ctrlo{1}    & \ctrl{1}     & \qw \\
      & \gate{R_y}   & \qw &   &   & \gate{R_y^1} & \gate{R_y^2} & \gate{R_y^3} & \gate{R_y^4} & \qw \\
      \\
      }
    \\
    \hline
      \tabincell{l}{\\GATEZ\\  3;  1,  2\\ -0.0073 -0.6907 -0.3105 -0.4417} 
    & $B_{1}^{3}$
    & \tabincell{l}{\\${{C}^{2}}{{R}_{z}}(3;1,2)$\\$2{{\phi }_{1}}=2\pi \cdot (-0.0073)$\\$2{{\phi }_{2}}=2\pi \cdot (-0.6907)$\\$2{{\phi }_{3}}=2\pi \cdot (-0.3105)$\\$2{{\phi }_{4}}=2\pi \cdot (-0.4417)$} 
    & \Qcircuit @C=0.4em @R=0.1em @!R{
      & \ctrlboth{1} & \qw &   &   & \ctrlo{1}    & \ctrlo{1}    & \ctrl{1}     & \ctrl{1}     & \qw \\
      & \ctrlboth{1} & \qw & = &   & \ctrlo{1}    & \ctrl{1}     & \ctrlo{1}    & \ctrl{1}     & \qw \\
      & \gate{R_z}   & \qw &   &   & \gate{R_z^1} & \gate{R_z^2} & \gate{R_z^3} & \gate{R_z^4} & \qw \\
      \\
      }
    \\
    \hline
\end{longtable}
\end{center}
\end{spacing}

\bigskip

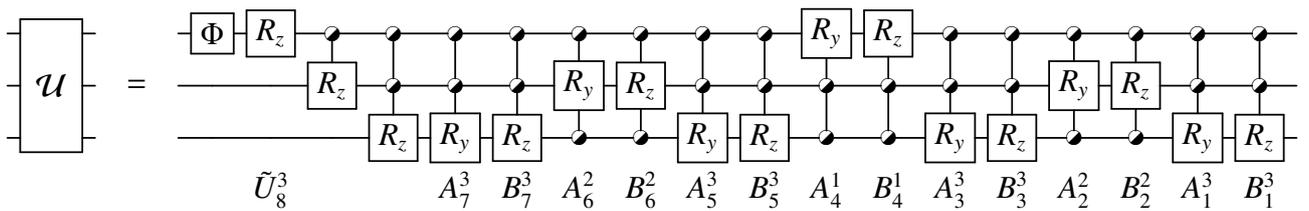
\begin{figure*}[!h]
\centering
\[
\Qcircuit @C=0.4em @R=0.1em @!R{
& \multigate{2}{\mathcal{U}} & \qw &                                          & & \gate{\Phi} & \gate{R_z}         & \ctrlboth{1} & \ctrlboth{1} & \ctrlboth{1} & \ctrlboth{1} & \ctrlboth{1} & \ctrlboth{1} & \ctrlboth{1} & \ctrlboth{1} & \gate{R_y}   & \gate{R_z}   & \ctrlboth{1} & \ctrlboth{1} & \ctrlboth{1} & \ctrlboth{1} & \ctrlboth{1} & \ctrlboth{1} & \qw \\
& \ghost{\mathcal{U}}        & \qw & \push{\rule{.3em}{0em}=\rule{.3em}{0em}} & & \qw         & \qw                & \gate{R_z}   & \ctrlboth{1} & \ctrlboth{1} & \ctrlboth{1} & \gate{R_y}   & \gate{R_z}   & \ctrlboth{1} & \ctrlboth{1} & \ctrlboth{-1}& \ctrlboth{-1}& \ctrlboth{1} & \ctrlboth{1} & \gate{R_y}   & \gate{R_z}   & \ctrlboth{1} & \ctrlboth{1} & \qw \\
& \ghost{\mathcal{U}}        & \qw &                                          & & \qw         & \qw                & \qw          & \gate{R_z}   & \gate{R_y}   & \gate{R_z}   & \ctrlboth{-1}& \ctrlboth{-1}& \gate{R_y}   & \gate{R_z}   & \ctrlboth{-1}& \ctrlboth{-1}& \gate{R_y}   & \gate{R_z}   & \ctrlboth{-1}& \ctrlboth{-1}& \gate{R_y}   & \gate{R_z}   & \qw \\
&                            &     &                                          & &             & \tilde{U}_{8}^{3}  &              &              & A_{7}^{3}    & B_{7}^{3}    & A_{6}^{2}    & B_{6}^{2}    & A_{5}^{3}    & B_{5}^{3}    & A_{4}^{1}    & B_{4}^{1}    & A_{3}^{3}    & B_{3}^{3}    & A_{2}^{2}    & B_{2}^{2}    & A_{1}^{3}    & B_{1}^{3}    &     \\
}
\]
\caption{\label{fig:example complexU}The circuit of $U$, the random 8-by-8 complex unitary matrix.}
\end{figure*}

The final quantum circuit of the complex unitary matrix $U$ given by Eq.~\ref{eq:random8by8matrix} is shown in Fig.~\ref{fig:example complexU}. 

It requires 8 gates for $\tilde{U}_{8}^{3}$, $4\times7$ $R_y$ gates for $A_{p}^{i}$, and $4\times7$ $R_z$ gates for $B_{p}^{i}$, thus 64 gates in total to build the equivalent quantum circuit.

\subsection{\label{sec:example realU}Real unitary matrix}

An 8-by-8 real unitary matrix $U$ is generated randomly as the following: 
\begin{equation}
U=\left( \begin{smallmatrix}
   -0.2991 & 0.1387 & 0.5667 & 0.0990 & 0.5322 & 0.1225 & 0.4933 & -0.1371  \\
   -0.3284 & -0.4431 & -0.0415 & -0.1607 & -0.4763 & -0.0154 & 0.3696 & -0.5519  \\
   -0.3766 & -0.7104 & 0.1156 & -0.1142 & 0.2958 & 0.0275 & -0.3785 & 0.3094  \\
   -0.3183 & 0.1523 & -0.7265 & -0.1032 & 0.5063 & -0.0310 & 0.0024 & -0.2826  \\
   -0.3415 & 0.0206 & -0.0599 & 0.9064 & -0.1500 & -0.1088 & -0.1481 & -0.0394  \\
   -0.4157 & 0.4240 & 0.2668 & -0.2504 & -0.1871 & 0.2838 & -0.5707 & -0.2689  \\
   -0.3908 & 0.1596 & -0.2301 & -0.0746 & -0.2757 & 0.4422 & 0.3534 & 0.6057  \\
   -0.3427 & 0.2257 & 0.0914 & -0.2255 & -0.1124 & -0.8337 & 0.0655 & 0.2456  \\
\end{smallmatrix} \right) .
\label{eq:random8by8realmatrix}
\end{equation}

Recursive CSD decomposition gives 
$ U=\tilde{A}_{1}^{3}\tilde{A}_{2}^{2}\tilde{A}_{3}^{3}\tilde{A}_{4}^{1}\tilde{A}_{5}^{3}\tilde{A}_{6}^{2}\tilde{A}_{7}^{3}\tilde{U}_{8}^{3} $. 
Their corresponding quantum gates are detailed in the output file produced by \emph{Qcompiler}, as explained below.

\begin{spacing}{1}
\begin{center}
\begin{longtable}{l|l|ll}
    \hline
      Output of the package &  Matrix & \multicolumn{2}{c}{Gate} \\ 
    \hline
    \endhead
      \tabincell{l}{\\GATEPI\\  1;\\  Y} 
    & 
    & \tabincell{l}{\\$\Pi (1;)$} 
    & \Qcircuit @C=0.4em @R=0.1em @!R{
      & \gate{\pi} & \qw \\
      & \qw        & \qw \\
      & \qw        & \qw \\
      \\
      }
    \\
      \tabincell{l}{\\GATEPI\\  2;  1\\  Y  N} 
    & $\tilde{U}_{8}^{3}$
    & \tabincell{l}{\\${{C}^{1}}\Pi (2;{1}_{0})$} 
    & \Qcircuit @C=0.4em @R=0.1em @!R{
      & \ctrlo{1}  & \qw \\
      & \gate{\pi} & \qw \\
      & \qw        & \qw \\
      \\
      }
    \\
      \tabincell{l}{\\GATEPI\\  3;  1,  2\\  Y  Y  N  Y} 
    & 
    & \tabincell{l}{\\${{C}^{2}}\Pi (3;{(1,2)}_{00})$\\${{C}^{2}}\Pi (3;{(1,2)}_{01})$\\${{C}^{2}}\Pi (3;{(1,2)}_{11})$} 
    & \Qcircuit @C=0.4em @R=0.1em @!R{
      & \ctrlo{1}  & \ctrlo{1}  & \ctrl{1}   & \qw \\
      & \ctrlo{1}  & \ctrl{1}   & \ctrl{1}   & \qw \\
      & \gate{\pi} & \gate{\pi} & \gate{\pi} & \qw \\
      \\
      }
    \\
    &
    & \multicolumn{2}{l}{
      \tabincell{l}{Note: This part specifies the subset of gates:\\
      \Qcircuit @C=0.4em @R=0.1em @!R{
      & \gate{\pi} & \ctrlboth{1} & \ctrlboth{1} & \qw \\
      & \qw        & \gate{\pi}   & \ctrlboth{1} & \qw \\
      & \qw        & \qw          & \gate{\pi}   & \qw \\
      }\\
      \smallskip}
    }
    \\
    \hline
      \tabincell{l}{\\GATEY\\  3;  1,  2\\ -0.3188  0.4501 -0.4725  0.2393} 
    & $\tilde{A}_{7}^{3}$
    & \tabincell{l}{\\${{C}^{2}}{{R}_{y}}(3;1,2)$\\$2{{\theta }_{1}}=2\pi \cdot (-0.3188)$\\$2{{\theta }_{2}}=2\pi \cdot 0.4501$\\$2{{\theta }_{3}}=2\pi \cdot (-0.4725)$\\$2{{\theta }_{4}}=2\pi \cdot 0.2393$}
    & \Qcircuit @C=0.4em @R=0.1em @!R{
      & \ctrlboth{1} & \qw &   &   & \ctrlo{1}    & \ctrlo{1}    & \ctrl{1}     & \ctrl{1}     & \qw \\
      & \ctrlboth{1} & \qw & = &   & \ctrlo{1}    & \ctrl{1}     & \ctrlo{1}    & \ctrl{1}     & \qw \\
      & \gate{R_y}   & \qw &   &   & \gate{R_y^1} & \gate{R_y^2} & \gate{R_y^3} & \gate{R_y^4} & \qw \\
      \\
      }
    \\
    \hline
      \tabincell{l}{\\GATEY\\  2;  1,  3\\  0.0624  0.4624  0.0128  0.4629} 
    & $\tilde{A}_{6}^{2}$
    & \tabincell{l}{\\${{C}^{2}}{{R}_{y}}(2;1,3)$\\$2{{\theta }_{1}}=2\pi \cdot 0.0624$\\$2{{\theta }_{2}}=2\pi \cdot 0.4624$\\$2{{\theta }_{3}}=2\pi \cdot 0.0128$\\$2{{\theta }_{4}}=2\pi \cdot 0.4629$}
    & \Qcircuit @C=0.4em @R=0.1em @!R{
      & \ctrlboth{1} & \qw &   &   & \ctrlo{1}    & \ctrlo{1}    & \ctrl{1}     & \ctrl{1}     & \qw \\
      & \gate{R_y}   & \qw & = &   & \gate{R_y^1} & \gate{R_y^2} & \gate{R_y^3} & \gate{R_y^4} & \qw \\
      & \ctrlboth{-1}& \qw &   &   & \ctrlo{-1}   & \ctrl{-1}    & \ctrlo{-1}   & \ctrl{-1}    & \qw \\
      \\
      }
    \\
    \hline
      \tabincell{l}{\\GATEY\\  3;  1,  2\\ -0.1026  0.4866 -0.0855 -0.2735} 
    & $\tilde{A}_{5}^{3}$
    & \tabincell{l}{\\${{C}^{2}}{{R}_{y}}(3;1,2)$\\$2{{\theta }_{1}}=2\pi \cdot (-0.1026)$\\$2{{\theta }_{2}}=2\pi \cdot 0.4866$\\$2{{\theta }_{3}}=2\pi \cdot (-0.0855)$\\$2{{\theta }_{4}}=2\pi \cdot (-0.2735)$}
    & \Qcircuit @C=0.4em @R=0.1em @!R{
      & \ctrlboth{1} & \qw &   &   & \ctrlo{1}    & \ctrlo{1}    & \ctrl{1}     & \ctrl{1}     & \qw \\
      & \ctrlboth{1} & \qw & = &   & \ctrlo{1}    & \ctrl{1}     & \ctrlo{1}    & \ctrl{1}     & \qw \\
      & \gate{R_y}   & \qw &   &   & \gate{R_y^1} & \gate{R_y^2} & \gate{R_y^3} & \gate{R_y^4} & \qw \\
      \\
      }
    \\
    \hline
      \tabincell{l}{\\GATEY\\  1;  2,  3\\ -0.0152  0.1125 -0.3422 -0.4830} 
    & $\tilde{A}_{4}^{1}$
    & \tabincell{l}{\\${{C}^{2}}{{R}_{y}}(1;2,3)$\\$2{{\theta }_{1}}=2\pi \cdot (-0.0152)$\\$2{{\theta }_{2}}=2\pi \cdot 0.1125$\\$2{{\theta }_{3}}=2\pi \cdot (-0.3422)$\\$2{{\theta }_{4}}=2\pi \cdot (-0.4830)$}
    & \Qcircuit @C=0.4em @R=0.1em @!R{
      & \gate{R_y}   & \qw &   &   & \gate{R_y^1} & \gate{R_y^2} & \gate{R_y^3} & \gate{R_y^4} & \qw \\
      & \ctrlboth{-1}& \qw & = &   & \ctrlo{-1}   & \ctrlo{-1}   & \ctrl{-1}    & \ctrl{-1}    & \qw \\
      & \ctrlboth{-1}& \qw &   &   & \ctrlo{-1}   & \ctrl{-1}    & \ctrlo{-1}   & \ctrl{-1}    & \qw \\
      \\
      }
    \\
    \hline
      \tabincell{l}{\\GATEY\\  3;  1,  2\\  0.3353 -0.3157 -0.2717  0.1060} 
    & $\tilde{A}_{3}^{3}$
    & \tabincell{l}{\\${{C}^{2}}{{R}_{y}}(3;1,2)$\\$2{{\theta }_{1}}=2\pi \cdot 0.3353$\\$2{{\theta }_{2}}=2\pi \cdot (-0.3157)$\\$2{{\theta }_{3}}=2\pi \cdot (-0.2717)$\\$2{{\theta }_{4}}=2\pi \cdot 0.1060$}
    & \Qcircuit @C=0.4em @R=0.1em @!R{
      & \ctrlboth{1} & \qw &   &   & \ctrlo{1}    & \ctrlo{1}    & \ctrl{1}     & \ctrl{1}     & \qw \\
      & \ctrlboth{1} & \qw & = &   & \ctrlo{1}    & \ctrl{1}     & \ctrlo{1}    & \ctrl{1}     & \qw \\
      & \gate{R_y}   & \qw &   &   & \gate{R_y^1} & \gate{R_y^2} & \gate{R_y^3} & \gate{R_y^4} & \qw \\
      \\
      }
    \\
    \hline
      \tabincell{l}{\\GATEY\\  2;  1,  3\\  0.2953  0.3166  0.2673  0.4561} 
    & $\tilde{A}_{2}^{2}$
    & \tabincell{l}{\\${{C}^{2}}{{R}_{y}}(2;1,3)$\\$2{{\theta }_{1}}=2\pi \cdot 0.2953$\\$2{{\theta }_{2}}=2\pi \cdot 0.3166$\\$2{{\theta }_{3}}=2\pi \cdot 0.2673$\\$2{{\theta }_{4}}=2\pi \cdot 0.4561$}
    & \Qcircuit @C=0.4em @R=0.1em @!R{
      & \ctrlboth{1} & \qw &   &   & \ctrlo{1}    & \ctrlo{1}    & \ctrl{1}     & \ctrl{1}     & \qw \\
      & \gate{R_y}   & \qw & = &   & \gate{R_y^1} & \gate{R_y^2} & \gate{R_y^3} & \gate{R_y^4} & \qw \\
      & \ctrlboth{-1}& \qw &   &   & \ctrlo{-1}   & \ctrl{-1}    & \ctrlo{-1}   & \ctrl{-1}    & \qw \\
      \\
      }
    \\
    \hline
      \tabincell{l}{\\GATEY\\  3;  1,  2\\  0.2999 -0.2617 -0.4684 -0.2811} 
    & $\tilde{A}_{1}^{3}$
    & \tabincell{l}{\\${{C}^{2}}{{R}_{y}}(3;1,2)$\\$2{{\theta }_{1}}=2\pi \cdot 0.2999$\\$2{{\theta }_{2}}=2\pi \cdot (-0.2617)$\\$2{{\theta }_{3}}=2\pi \cdot (-0.4684)$\\$2{{\theta }_{4}}=2\pi \cdot (-0.2811)$}
    & \Qcircuit @C=0.4em @R=0.1em @!R{
      & \ctrlboth{1} & \qw &   &   & \ctrlo{1}    & \ctrlo{1}    & \ctrl{1}     & \ctrl{1}     & \qw \\
      & \ctrlboth{1} & \qw & = &   & \ctrlo{1}    & \ctrl{1}     & \ctrlo{1}    & \ctrl{1}     & \qw \\
      & \gate{R_y}   & \qw &   &   & \gate{R_y^1} & \gate{R_y^2} & \gate{R_y^3} & \gate{R_y^4} & \qw \\
      \\
      }
    \\
    \hline
\end{longtable}
\end{center}
\end{spacing}

\bigskip

\begin{figure*}[!h]
\centering
\[
\Qcircuit @C=0.4em @R=0.1em @!R{
& \multigate{2}{\mathcal{U}} & \qw &                                          & & \gate{\pi} & \ctrlo{1}  & \ctrlo{1}          & \ctrlo{1}  & \ctrl{1}   & \ctrlboth{1}      & \ctrlboth{1}      & \ctrlboth{1}      & \gate{R_y}        & \ctrlboth{1}      & \ctrlboth{1}      & \ctrlboth{1}      & \qw \\
& \ghost{\mathcal{U}}        & \qw & \push{\rule{.3em}{0em}=\rule{.3em}{0em}} & & \qw        & \gate{\pi} & \ctrlo{1}          & \ctrl{1}   & \ctrl{1}   & \ctrlboth{1}      & \gate{R_y}        & \ctrlboth{1}      & \ctrlboth{-1}     & \ctrlboth{1}      & \gate{R_y}        & \ctrlboth{1}      & \qw \\
& \ghost{\mathcal{U}}        & \qw &                                          & & \qw        & \qw        & \gate{\pi}         & \gate{\pi} & \gate{\pi} & \gate{R_y}        & \ctrlboth{-1}     & \gate{R_y}        & \ctrlboth{-1}     & \gate{R_y}        & \ctrlboth{-1}     & \gate{R_y}        & \qw \\
&                            &     &                                          & &            &            & \tilde{U}_{8}^{3}  &            &            & \tilde{A}_{7}^{3} & \tilde{A}_{6}^{2} & \tilde{A}_{5}^{3} & \tilde{A}_{4}^{1} & \tilde{A}_{3}^{3} & \tilde{A}_{2}^{2} & \tilde{A}_{1}^{3} &     \\
}
\]
\caption{\label{fig:example realU}The circuit of $U$, the random 8-by-8 real unitary matrix.}
\end{figure*}
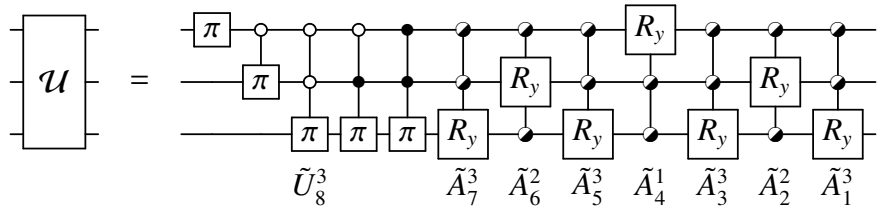

The final quantum circuit of the real unitary matrix $U$ given by Eq.~\ref{eq:random8by8realmatrix} is shown in Fig.~\ref{fig:example realU}, which is significantly simpler than that shown in Fig.~\ref{fig:example complexU} for a complex unitary matrix of the same size.

It requires 5 $\Pi$ gates for $\tilde{U}_{8}^{3}$, and $4\times7$ $R_y$ gates for $\tilde{A}_{p}^{i}$, thus 33 gates in total to build the equivalent quantum circuit, significantly simpler than its complex counterpart.

\subsection{\label{sec:example square}Random walk on a square graph}

Quantum walks have recently been explored for their non-intuitive dynamics, which may hold the key to radically new quantum algorithms \cite{Kempe2003, Douglas2008, Berry2011}.  Douglas and Wang \cite{Douglas2009} recently presented efficient quantum circuits for several families of highly symmetrical graphs.  Here we use \emph{Qcompiler} to build quantum circuits for quantum walk on general graphs. 

The walker's quantum state is given : 
\begin{equation}
\left| \psi  \right\rangle =\sum\limits_{i=1}^{N}{\sum\limits_{j\in S}{{{A}_{i,j}}\left| i,j \right\rangle }},
\end{equation}
where $N$ is the number of nodes of the given graph, $\left| i \right\rangle $ represents the $i$th node state, 
$\left| j \right\rangle $ the coin state which connects the $i$th and $j$th node, and $S$ is determined by the graph adjacent matrix.

To realize one step quantum walk, we first apply a coin operator $\hat{C}=\underset{i=1,...,N}{\mathop{diag}}\,({{\hat{C}}_{i}})$ on the walker, where
\begin{equation}
\hat{C}_{i}=\sum\limits_{j\in S}{\sum\limits_{k\in S}{c_{j,k}^{i}\left| i,k \right\rangle \left\langle  i,j \right|}} .
\end{equation}
In this work, we choose the Grover Coin  \cite{Shenvi2003} so that the quantum walk can be represented by a real matrix.
We then apply the translation operator 
\begin{equation}
\hat{T}=\sum\limits_{i=1}^{N}{\sum\limits_{j\in S}{\left| j,i \right\rangle \left\langle  i,j \right|}}.
\end{equation}
The unitary matrix for the complete step is
$\hat{U}=\hat{T}\hat{C}$.

\begin{figure}[!h]
\centering
\includegraphics[scale=0.8]{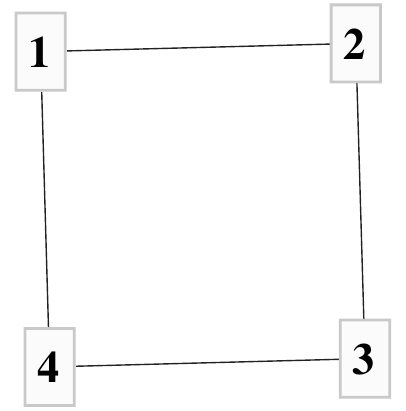}
\caption{\label{fig:square}The square graph}
\end{figure}

For a simple square graph, shown in Fig.~\ref{fig:square}, the quantum walk operator is
\[
\hat{U}=\hat{T}\hat{C}=
\left( \begin{smallmatrix}
   {} & {} & 1 & {} & {} & {} & {} & {}  \\
   {} & {} & {} & {} & {} & {} & 1 & {}  \\
   1 & {} & {} & {} & {} & {} & {} & {}  \\
   {} & {} & {} & {} & 1 & {} & {} & {}  \\
   {} & {} & {} & 1 & {} & {} & {} & {}  \\
   {} & {} & {} & {} & {} & {} & {} & 1  \\
   {} & 1 & {} & {} & {} & {} & {} & {}  \\
   {} & {} & {} & {} & {} & 1 & {} & {}  \\
\end{smallmatrix} \right)
\left( \begin{smallmatrix}
   0 & 1 & {} & {} & {} & {} & {} & {}  \\
   1 & 0 & {} & {} & {} & {} & {} & {}  \\
   {} & {} & 0 & 1 & {} & {} & {} & {}  \\
   {} & {} & 1 & 0 & {} & {} & {} & {}  \\
   {} & {} & {} & {} & 0 & 1 & {} & {}  \\
   {} & {} & {} & {} & 1 & 0 & {} & {}  \\
   {} & {} & {} & {} & {} & {} & 0 & 1  \\
   {} & {} & {} & {} & {} & {} & 1 & 0  \\
\end{smallmatrix} \right)
=\left( \begin{smallmatrix}
   0 & 0 & 0 & 1 & 0 & 0 & 0 & 0  \\
   0 & 0 & 0 & 0 & 0 & 0 & 0 & 1  \\
   0 & 1 & 0 & 0 & 0 & 0 & 0 & 0  \\
   0 & 0 & 0 & 0 & 0 & 1 & 0 & 0  \\
   0 & 0 & 1 & 0 & 0 & 0 & 0 & 0  \\
   0 & 0 & 0 & 0 & 0 & 0 & 1 & 0  \\
   1 & 0 & 0 & 0 & 0 & 0 & 0 & 0  \\
   0 & 0 & 0 & 0 & 1 & 0 & 0 & 0  \\
\end{smallmatrix} \right)
\]

The output file of Qcompiler is given as the following with its quantum circuit shown in Fig.~\ref{fig:example square}.

\begin{spacing}{1}
\begin{center}
\begin{longtable}{l|l|ll}
    \hline
      Output of the package &  Matrix & \multicolumn{2}{c}{Gate} \\ 
    \hline
    \endhead
      \tabincell{l}{\\GATEPI\\  1;\\  Y} 
    & 
    & \tabincell{l}{\\$\Pi (1;)$} 
    & \Qcircuit @C=0.4em @R=0.1em @!R{
      & \gate{\pi} & \qw \\
      & \qw        & \qw \\
      & \qw        & \qw \\
      \\
      }
    \\
      \tabincell{l}{\\GATEPI\\  2;  1\\  Y  N} 
    & $\tilde{U}_{8}^{3}$
    & \tabincell{l}{\\${{C}^{1}}\Pi (2;{1}_{0})$} 
    & \Qcircuit @C=0.4em @R=0.1em @!R{
      & \ctrlo{1}  & \qw \\
      & \gate{\pi} & \qw \\
      & \qw        & \qw \\
      \\
      }
    \\
      \tabincell{l}{\\GATEPI\\  3;  1,  2\\  N  Y  Y  N} 
    & 
    & \tabincell{l}{\\${{C}^{2}}\Pi (3;{(1,2)}_{01})$\\${{C}^{2}}\Pi (3;{(1,2)}_{10})$} 
    & \Qcircuit @C=0.4em @R=0.1em @!R{
      & \ctrlo{1}  & \ctrl{1}   & \qw \\
      & \ctrl{1}   & \ctrlo{1}  & \qw \\
      & \gate{\pi} & \gate{\pi} & \qw \\
      \\
      }
    \\
    &
    & \multicolumn{2}{l}{
      \tabincell{l}{Note: This part specifies the subset of gates:\\
      \Qcircuit @C=0.4em @R=0.1em @!R{
      & \gate{\pi} & \ctrlboth{1} & \ctrlboth{1} & \qw \\
      & \qw        & \gate{\pi}   & \ctrlboth{1} & \qw \\
      & \qw        & \qw          & \gate{\pi}   & \qw \\
      }\\
      \smallskip}
    }
    \\
    \hline
      \tabincell{l}{\\GATEY\\  3;  1,  2\\  0.5000  0.0000  0.0000 -0.5000} 
    & $\tilde{A}_{7}^{3}$
    & \tabincell{l}{\\${{C}^{2}}{{R}_{y}}(3;1,2)$\\$2{{\theta }_{1}}=2\pi \cdot 0.5=\pi$\\$2{{\theta }_{4}}=2\pi \cdot (-0.5)=-\pi$}
    & \Qcircuit @C=0.4em @R=0.1em @!R{
      & \ctrlo{1}    & \ctrl{1}     & \qw \\
      & \ctrlo{1}    & \ctrl{1}     & \qw \\
      & \gate{R_y^1} & \gate{R_y^4} & \qw \\
      \\
      }
    \\
    \hline
      \tabincell{l}{\\GATEY\\  2;  1,  3\\  0.0000  0.5000  0.0000  0.5000} 
    & $\tilde{A}_{6}^{2}$
    & \tabincell{l}{\\${{C}^{2}}{{R}_{y}}(2;1,3)$\\$2{{\theta }_{2}}=2\pi \cdot 0.5=\pi$\\$2{{\theta }_{4}}=2\pi \cdot 0.5=\pi$}
    & \Qcircuit @C=0.4em @R=0.1em @!R{
      & \ctrlo{1}    & \ctrl{1}     & \qw \\
      & \gate{R_y^2} & \gate{R_y^4} & \qw \\
      & \ctrl{-1}    & \ctrl{-1}    & \qw \\
      \\
      }
    \\
    \hline
      \tabincell{l}{\\GATEY\\  3;  1,  2\\ -0.5000 -0.5000  0.0000  0.0000} 
    & $\tilde{A}_{5}^{3}$
    & \tabincell{l}{\\${{C}^{2}}{{R}_{y}}(3;1,2)$\\$2{{\theta }_{1}}=2\pi \cdot (-0.5)=-\pi$\\$2{{\theta }_{2}}=2\pi \cdot (-0.5)=-\pi$}
    & \Qcircuit @C=0.4em @R=0.1em @!R{
      & \ctrlo{1}    & \ctrlo{1}    & \qw \\
      & \ctrlo{1}    & \ctrl{1}     & \qw \\
      & \gate{R_y^1} & \gate{R_y^2} & \qw \\
      \\
      }
    \\
    \hline
      \tabincell{l}{\\GATEY\\  1;  2,  3\\  0.0000  0.0000 -0.5000  0.5000} 
    & $\tilde{A}_{4}^{1}$
    & \tabincell{l}{\\${{C}^{2}}{{R}_{y}}(1;2,3)$\\$2{{\theta }_{3}}=2\pi \cdot (-0.5)=-\pi$\\$2{{\theta }_{4}}=2\pi \cdot 0.5=\pi$}
    & \Qcircuit @C=0.4em @R=0.1em @!R{
      & \gate{R_y^3} & \gate{R_y^4} & \qw \\
      & \ctrl{-1}    & \ctrl{-1}    & \qw \\
      & \ctrlo{-1}   & \ctrl{-1}    & \qw \\
      \\
      }
    \\
    \hline
      \tabincell{l}{\\GATEY\\  3;  1,  2\\  0.0000  0.5000  0.5000  0.0000} 
    & $\tilde{A}_{3}^{3}$
    & \tabincell{l}{\\${{C}^{2}}{{R}_{y}}(3;1,2)$\\$2{{\theta }_{2}}=2\pi \cdot 0.5=\pi$\\$2{{\theta }_{3}}=2\pi \cdot 0.5=\pi$}
    & \Qcircuit @C=0.4em @R=0.1em @!R{
      & \ctrlo{1}    & \ctrl{1}     & \qw \\
      & \ctrl{1}     & \ctrlo{1}    & \qw \\
      & \gate{R_y^2} & \gate{R_y^3} & \qw \\
      \\
      }
    \\
    \hline
      \tabincell{l}{\\GATEY\\  2;  1,  3\\  0.0000 -0.5000  0.0000  0.5000} 
    & $\tilde{A}_{2}^{2}$
    & \tabincell{l}{\\${{C}^{2}}{{R}_{y}}(2;1,3)$\\$2{{\theta }_{2}}=2\pi \cdot (-0.5)=-\pi$\\$2{{\theta }_{4}}=2\pi \cdot 0.5=\pi$}
    & \Qcircuit @C=0.4em @R=0.1em @!R{
      & \ctrlo{1}    & \ctrl{1}     & \qw \\
      & \gate{R_y^2} & \gate{R_y^4} & \qw \\
      & \ctrl{-1}    & \ctrl{-1}    & \qw \\
      \\
      }
    \\
    \hline
      \tabincell{l}{\\GATEY\\  3;  1,  2\\  0.0000 -0.5000 -0.5000  0.0000} 
    & $\tilde{A}_{1}^{3}$
    & \tabincell{l}{\\${{C}^{2}}{{R}_{y}}(3;1,2)$\\$2{{\theta }_{2}}=2\pi \cdot (-0.5)=-\pi$\\$2{{\theta }_{3}}=2\pi \cdot (-0.5)=-\pi$}
    & \Qcircuit @C=0.4em @R=0.1em @!R{
      & \ctrlo{1}    & \ctrl{1}     & \qw \\
      & \ctrl{1}     & \ctrlo{1}    & \qw \\
      & \gate{R_y^2} & \gate{R_y^3} & \qw \\
      \\
      }
    \\
    \hline
\end{longtable}
\end{center}
\end{spacing}

\bigskip

\begin{center}
\begin{figure*}[!h]
\centering
\[
\Qcircuit @C=0.4em @R=0.1em @!R{
& \multigate{2}{\mathcal{U}} & \qw &                                          & & \gate{\pi} & \ctrlo{1}          & \ctrlo{1}  & \ctrl{1}   & \ctrlo{1}    & \ctrl{1}     & \ctrlo{1}    & \ctrl{1}     & \ctrlo{1}    & \ctrlo{1}    & \gate{R_y^3} & \gate{R_y^4} & \ctrlo{1}    & \ctrl{1}     & \ctrlo{1}    & \ctrl{1}     & \ctrlo{1}    & \ctrl{1}     & \qw \\
& \ghost{\mathcal{U}}        & \qw & \push{\rule{.3em}{0em}=\rule{.3em}{0em}} & & \qw        & \gate{\pi}         & \ctrl{1}   & \ctrlo{1}  & \ctrlo{1}    & \ctrl{1}     & \gate{R_y^2} & \gate{R_y^4} & \ctrlo{1}    & \ctrl{1}     & \ctrl{-1}    & \ctrl{-1}    & \ctrl{1}     & \ctrlo{1}    & \gate{R_y^2} & \gate{R_y^4} & \ctrl{1}     & \ctrlo{1}    & \qw \\
& \ghost{\mathcal{U}}        & \qw &                                          & & \qw        & \qw                & \gate{\pi} & \gate{\pi} & \gate{R_y^1} & \gate{R_y^4} & \ctrl{-1}    & \ctrl{-1}    & \gate{R_y^1} & \gate{R_y^2} & \ctrlo{-1}   & \ctrl{-1}    & \gate{R_y^2} & \gate{R_y^3} & \ctrl{-1}    & \ctrl{-1}    & \gate{R_y^2} & \gate{R_y^3} & \qw \\
&                            &     &                                          & &            & \tilde{U}_{8}^{3}  &            &            &         & \tilde{A}_{7}^{3} &         & \tilde{A}_{6}^{2} &         & \tilde{A}_{5}^{3} &         & \tilde{A}_{4}^{1} &         & \tilde{A}_{3}^{3} &         & \tilde{A}_{2}^{2} &         & \tilde{A}_{1}^{3} &     \\
}
\]
\caption{\label{fig:example square}The circuit of $U$, the random walk evolution on the square graph (Fig.~\ref{fig:square}).}
\end{figure*}
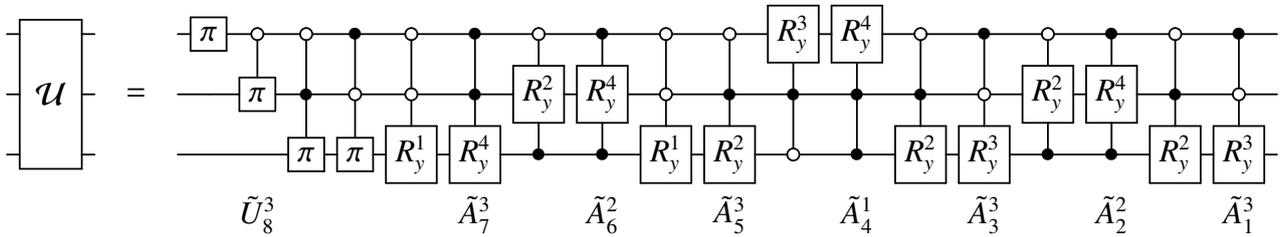
\end{center}

Compare to the random 8-by-8 real matrix given by Eq.~\ref{eq:random8by8realmatrix}, which requires 33 gates, the circuit of the square graph consists of only 18 gates, reflecting the simple symmetry of this graph.

\subsection{\label{sec:example star}Random walk on the 8-star graph}

\begin{figure}[!h]
\centering
\includegraphics[scale=0.8]{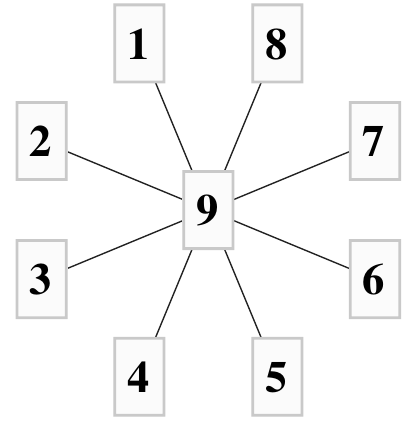}
\caption{\label{fig:star}The 8-star graph}
\end{figure}

Fig.~\ref{fig:star} shows another simple graph, namely the 8-star graph

For this graph, the quantum walk operator is 
\[
\begin{split}
\hat{U}=&\hat{T}\hat{C} \\
=&\left( \begin{smallmatrix}
   {} & {} & {} & {} & {} & {} & {} & {} & -0.75 & 0.25 & 0.25 & 0.25 & 0.25 & 0.25 & 0.25 & 0.25  \\
   {} & {} & {} & {} & {} & {} & {} & {} & 0.25 & -0.75 & 0.25 & 0.25 & 0.25 & 0.25 & 0.25 & 0.25  \\
   {} & {} & {} & {} & {} & {} & {} & {} & 0.25 & 0.25 & -0.75 & 0.25 & 0.25 & 0.25 & 0.25 & 0.25  \\
   {} & {} & {} & {} & {} & {} & {} & {} & 0.25 & 0.25 & 0.25 & -0.75 & 0.25 & 0.25 & 0.25 & 0.25  \\
   {} & {} & {} & {} & {} & {} & {} & {} & 0.25 & 0.25 & 0.25 & 0.25 & -0.75 & 0.25 & 0.25 & 0.25  \\
   {} & {} & {} & {} & {} & {} & {} & {} & 0.25 & 0.25 & 0.25 & 0.25 & 0.25 & -0.75 & 0.25 & 0.25  \\
   {} & {} & {} & {} & {} & {} & {} & {} & 0.25 & 0.25 & 0.25 & 0.25 & 0.25 & 0.25 & -0.75 & 0.25  \\
   {} & {} & {} & {} & {} & {} & {} & {} & 0.25 & 0.25 & 0.25 & 0.25 & 0.25 & 0.25 & 0.25 & -0.75  \\
   1 & {} & {} & {} & {} & {} & {} & {} & {} & {} & {} & {} & {} & {} & {} & {}  \\
   {} & 1 & {} & {} & {} & {} & {} & {} & {} & {} & {} & {} & {} & {} & {} & {}  \\
   {} & {} & 1 & {} & {} & {} & {} & {} & {} & {} & {} & {} & {} & {} & {} & {}  \\
   {} & {} & {} & 1 & {} & {} & {} & {} & {} & {} & {} & {} & {} & {} & {} & {}  \\
   {} & {} & {} & {} & 1 & {} & {} & {} & {} & {} & {} & {} & {} & {} & {} & {}  \\
   {} & {} & {} & {} & {} & 1 & {} & {} & {} & {} & {} & {} & {} & {} & {} & {}  \\
   {} & {} & {} & {} & {} & {} & 1 & {} & {} & {} & {} & {} & {} & {} & {} & {}  \\
   {} & {} & {} & {} & {} & {} & {} & 1 & {} & {} & {} & {} & {} & {} & {} & {}  \\
\end{smallmatrix} \right),
\end{split}
\]
which is a $16\times 16$ unitary matrix. The complete result is displayed below.
Note that since all of the rotation angles of $\tilde{A}_{1}^{4}$, $\tilde{A}_{2}^{3}$, $\tilde{A}_{3}^{4}$, $\tilde{A}_{4}^{2}$, $\tilde{A}_{5}^{4}$, $\tilde{A}_{6}^{3}$, $\tilde{A}_{7}^{4}$ are zeros, these gates vanish and are omitted here.

\begin{spacing}{1}
\begin{center}
\begin{longtable}{l|l|ll}
    \hline
      Output of the package &  Matrix & \multicolumn{2}{c}{Gate} \\ 
    \hline
    \endhead
      \tabincell{l}{\\GATEPI\\  1;\\  N} 
    & 
    & 
    & 
    \\
      \tabincell{l}{\\GATEPI\\  2;  1\\  N  N} 
    & 
    & 
    & 
    \\
      \tabincell{l}{\\GATEPI\\  3;  1,  2\\  N  N  N  N} 
    & 
    & 
    & 
    \\
      \tabincell{l}{\\GATEPI\\  4;  1,  2,  3\\  N  N  N  N  N  N  N  Y} 
    & $\tilde{U}_{16}^{4}$
    & \tabincell{l}{\\${{C}^{3}}\Pi (4;{(1,2,3)}_{111})$} 
    & \Qcircuit @C=0.4em @R=0.1em @!R{
      & \ctrl{1}   & \qw \\
      & \ctrl{1}   & \qw \\
      & \ctrl{1}   & \qw \\
      & \gate{\pi} & \qw \\
      \\
      }
    \\
    &
    & \multicolumn{2}{l}{
      \tabincell{l}{Note: This part specifies the subset of gates:\\
      \Qcircuit @C=0.4em @R=0.1em @!R{
      & \gate{\pi} & \ctrlboth{1} & \ctrlboth{1} & \ctrlboth{1} & \qw \\
      & \qw        & \gate{\pi}   & \ctrlboth{1} & \ctrlboth{1} & \qw \\
      & \qw        & \qw          & \gate{\pi}   & \ctrlboth{1} & \qw \\
      & \qw        & \qw          & \qw          & \gate{\pi}   & \qw \\
      }\\
      \smallskip}
    }
    \\
    \hline
      \tabincell{l}{\\GATEY\\  4;  1,  2,  3\\  0.0000  0.0000  0.0000  0.0000\\ -0.1153  0.1573 -0.3073 -0.0372} 
    & $\tilde{A}_{15}^{4}$
    & \tabincell{l}{\\${{C}^{3}}{{R}_{y}}(4;1,2,3)$\\$2{{\theta }_{5}}=2\pi \cdot (-0.1153)$\\$2{{\theta }_{6}}=2\pi \cdot 0.1573$\\$2{{\theta }_{7}}=2\pi \cdot (-0.3073)$\\$2{{\theta }_{8}}=2\pi \cdot (-0.0372)$}
    & \Qcircuit @C=0.4em @R=0.1em @!R{
      & \ctrl{1}     & \ctrl{1}     & \ctrl{1}     & \ctrl{1}     & \qw \\
      & \ctrlo{1}    & \ctrlo{1}    & \ctrl{1}     & \ctrl{1}     & \qw \\
      & \ctrlo{1}    & \ctrl{1}     & \ctrlo{1}    & \ctrl{1}     & \qw \\
      & \gate{R_y^5} & \gate{R_y^6} & \gate{R_y^7} & \gate{R_y^8} & \qw \\
      \\
      }
    \\
    \hline
      \tabincell{l}{\\GATEY\\  3;  1,  2,  4\\  0.0000  0.0000  0.0000  0.0000\\ -0.1289 -0.4029  0.0714  0.3210} 
    & $\tilde{A}_{14}^{3}$
    & \tabincell{l}{\\${{C}^{3}}{{R}_{y}}(3;1,2,4)$\\$2{{\theta }_{5}}=2\pi \cdot (-0.1289)$\\$2{{\theta }_{6}}=2\pi \cdot (-0.4029)$\\$2{{\theta }_{7}}=2\pi \cdot 0.0714$\\$2{{\theta }_{8}}=2\pi \cdot 0.3210$}
    & \Qcircuit @C=0.4em @R=0.1em @!R{
      & \ctrl{1}     & \ctrl{1}     & \ctrl{1}     & \ctrl{1}     & \qw \\
      & \ctrlo{1}    & \ctrlo{1}    & \ctrl{1}     & \ctrl{1}     & \qw \\
      & \gate{R_y^5} & \gate{R_y^6} & \gate{R_y^7} & \gate{R_y^8} & \qw \\
      & \ctrlo{-1}   & \ctrl{-1}    & \ctrlo{-1}   & \ctrl{-1}    & \qw \\
      \\
      }
    \\
    \hline
      \tabincell{l}{\\GATEY\\  4;  1,  2,  3\\  0.0000  0.0000  0.0000  0.0000\\  0.0392 -0.2637 -0.4685  0.1927} 
    & $\tilde{A}_{13}^{4}$
    & \tabincell{l}{\\${{C}^{3}}{{R}_{y}}(4;1,2,3)$\\$2{{\theta }_{5}}=2\pi \cdot 0.0392$\\$2{{\theta }_{6}}=2\pi \cdot (-0.2637)$\\$2{{\theta }_{7}}=2\pi \cdot (-0.4685)$\\$2{{\theta }_{8}}=2\pi \cdot 0.1927$}
    & \Qcircuit @C=0.4em @R=0.1em @!R{
      & \ctrl{1}     & \ctrl{1}     & \ctrl{1}     & \ctrl{1}     & \qw \\
      & \ctrlo{1}    & \ctrlo{1}    & \ctrl{1}     & \ctrl{1}     & \qw \\
      & \ctrlo{1}    & \ctrl{1}     & \ctrlo{1}    & \ctrl{1}     & \qw \\
      & \gate{R_y^5} & \gate{R_y^6} & \gate{R_y^7} & \gate{R_y^8} & \qw \\
      \\
      }
    \\
    \hline
      \tabincell{l}{\\GATEY\\  2;  1,  3,  4\\  0.0000  0.0000  0.0000  0.0000\\  0.0000  0.0000  0.0000  0.5000} 
    & $\tilde{A}_{12}^{2}$
    & \tabincell{l}{\\${{C}^{3}}{{R}_{y}}(2;1,3,4)$\\$2{{\theta }_{8}}=2\pi \cdot 0.5=\pi$}
    & \Qcircuit @C=0.4em @R=0.1em @!R{
      & \ctrl{1}     & \qw \\
      & \gate{R_y^8} & \qw \\
      & \ctrl{-1}    & \qw \\
      & \ctrl{-1}    & \qw \\
      \\
      }
    \\
    \hline
      \tabincell{l}{\\GATEY\\  4;  1,  2,  3\\  0.0000  0.0000  0.0000  0.0000\\ -0.0392  0.2637  0.4685  0.1927} 
    & $\tilde{A}_{11}^{4}$
    & \tabincell{l}{\\${{C}^{3}}{{R}_{y}}(4;1,2,3)$\\$2{{\theta }_{5}}=2\pi \cdot (-0.0392)$\\$2{{\theta }_{6}}=2\pi \cdot 0.2637$\\$2{{\theta }_{7}}=2\pi \cdot 0.4685$\\$2{{\theta }_{8}}=2\pi \cdot 0.1927$}
    & \Qcircuit @C=0.4em @R=0.1em @!R{
      & \ctrl{1}     & \ctrl{1}     & \ctrl{1}     & \ctrl{1}     & \qw \\
      & \ctrlo{1}    & \ctrlo{1}    & \ctrl{1}     & \ctrl{1}     & \qw \\
      & \ctrlo{1}    & \ctrl{1}     & \ctrlo{1}    & \ctrl{1}     & \qw \\
      & \gate{R_y^5} & \gate{R_y^6} & \gate{R_y^7} & \gate{R_y^8} & \qw \\
      \\
      }
    \\
    \hline
      \tabincell{l}{\\GATEY\\  3;  1,  2,  4\\  0.0000  0.0000  0.0000  0.0000\\  0.1289  0.4029 -0.0714  0.3210} 
    & $\tilde{A}_{10}^{3}$
    & \tabincell{l}{\\${{C}^{3}}{{R}_{y}}(3;1,2,4)$\\$2{{\theta }_{5}}=2\pi \cdot 0.1289$\\$2{{\theta }_{6}}=2\pi \cdot 0.4029$\\$2{{\theta }_{7}}=2\pi \cdot (-0.0714)$\\$2{{\theta }_{8}}=2\pi \cdot 0.3210$}
    & \Qcircuit @C=0.4em @R=0.1em @!R{
      & \ctrl{1}     & \ctrl{1}     & \ctrl{1}     & \ctrl{1}     & \qw \\
      & \ctrlo{1}    & \ctrlo{1}    & \ctrl{1}     & \ctrl{1}     & \qw \\
      & \gate{R_y^5} & \gate{R_y^6} & \gate{R_y^7} & \gate{R_y^8} & \qw \\
      & \ctrlo{-1}   & \ctrl{-1}    & \ctrlo{-1}   & \ctrl{-1}    & \qw \\
      \\
      }
    \\
    \hline
      \tabincell{l}{\\GATEY\\  4;  1,  2,  3\\  0.0000  0.0000  0.0000  0.0000\\  0.1153 -0.1573  0.3073 -0.0372} 
    & $\tilde{A}_{9}^{4}$
    & \tabincell{l}{\\${{C}^{3}}{{R}_{y}}(4;1,2,3)$\\$2{{\theta }_{5}}=2\pi \cdot 0.1153$\\$2{{\theta }_{6}}=2\pi \cdot (-0.1573)$\\$2{{\theta }_{7}}=2\pi \cdot 0.3073$\\$2{{\theta }_{8}}=2\pi \cdot (-0.0372)$}
    & \Qcircuit @C=0.4em @R=0.1em @!R{
      & \ctrl{1}     & \ctrl{1}     & \ctrl{1}     & \ctrl{1}     & \qw \\
      & \ctrlo{1}    & \ctrlo{1}    & \ctrl{1}     & \ctrl{1}     & \qw \\
      & \ctrlo{1}    & \ctrl{1}     & \ctrlo{1}    & \ctrl{1}     & \qw \\
      & \gate{R_y^5} & \gate{R_y^6} & \gate{R_y^7} & \gate{R_y^8} & \qw \\
      \\
      }
    \\
    \hline
      \tabincell{l}{\\GATEY\\  1;  2,  3,  4\\ -0.5000 -0.5000 -0.5000 -0.5000\\ -0.5000 -0.5000 -0.5000 -0.5000} 
    & $\tilde{A}_{8}^{1}$
    & \tabincell{l}{\\${{C}^{3}}{{R}_{y}}(1;2,3,4)$\\$2{{\theta }_{1}}=2{{\theta }_{2}}=2{{\theta }_{3}}=2{{\theta }_{4}}=$\\$2{{\theta }_{5}}=2{{\theta }_{6}}=2{{\theta }_{7}}=2{{\theta }_{8}}=$\\$2\pi \cdot (-0.5)=-\pi$}
    & \Qcircuit @C=0.4em @R=0.1em @!R{
      & \gate{R_y}    & \qw &   & & \gate{R_y^1}  & \gate{R_y^2}  & \qw &                                            & & \gate{R_y^8}  & \qw \\
      & \ctrlboth{-1} & \qw & = & & \ctrlo{-1}    & \ctrlo{-1}    & \qw & \push{\rule{.3em}{0em}...\rule{.3em}{0em}} & & \ctrl{-1}     & \qw \\
      & \ctrlboth{-1} & \qw &   & & \ctrlo{-1}    & \ctrlo{-1}    & \qw &                                            & & \ctrl{-1}     & \qw \\
      & \ctrlboth{-1} & \qw &   & & \ctrlo{-1}    & \ctrl{-1}     & \qw &                                            & & \ctrl{-1}     & \qw \\
      \\
      }
    \\
    \hline
\end{longtable}
\end{center}
\end{spacing}

Putting these gates together, we obtain its quantum circuit as shown in Fig.~\ref{fig:example star}.

\begin{center}
\begin{figure*}[!h]
\centering
\[
 \Qcircuit @C=0.4em @R=0.1em @!R{
& \multigate{3}{\mathcal{U}} & \qw &                                          & & \ctrl{1}           & \ctrl{1}     & \ctrl{1}     & \ctrl{1}     & \ctrl{1}     & \ctrl{1}     & \ctrl{1}     & \ctrl{1}     & \ctrl{1}     & \ctrl{1}     & \ctrl{1}     & \ctrl{1}     & \ctrl{1}     & \ctrl{1}           & \qw \\       
& \ghost{\mathcal{U}}        & \qw & \push{\rule{.3em}{0em}=\rule{.3em}{0em}} & & \ctrl{1}           & \ctrlo{1}    & \ctrlo{1}    & \ctrl{1}     & \ctrl{1}     & \ctrlo{1}    & \ctrlo{1}    & \ctrl{1}     & \ctrl{1}     & \ctrlo{1}    & \ctrlo{1}    & \ctrl{1}     & \ctrl{1}     & \gate{R_y}         & \qw \\       
& \ghost{\mathcal{U}}        & \qw &                                          & & \ctrl{1}           & \ctrlo{1}    & \ctrl{1}     & \ctrlo{1}    & \ctrl{1}     & \gate{R_y}   & \gate{R_y}   & \gate{R_y}   & \gate{R_y}   & \ctrlo{1}    & \ctrl{1}     & \ctrlo{1}    & \ctrl{1}     & \ctrl{-1}          & \qw \\       
& \ghost{\mathcal{U}}        & \qw &                                          & & \gate{\pi}         & \gate{R_y}   & \gate{R_y}   & \gate{R_y}   & \gate{R_y}   & \ctrlo{-1}   & \ctrl{-1}    & \ctrlo{-1}   & \ctrl{-1}    & \gate{R_y}   & \gate{R_y}   & \gate{R_y}   & \gate{R_y}   & \ctrl{-1}          & \qw \\       
&                            &     &                                          & & \tilde{U}_{16}^{4} &              & \tilde{A}_{15}^{4} &        &              &              & \tilde{A}_{14}^{3} &        &              &              & \tilde{A}_{13}^{4} &        &              & \tilde{A}_{12}^{2} &     \\       
& \ctrl{1}     & \ctrl{1}     & \ctrl{1}     & \ctrl{1}     & \ctrl{1}     & \ctrl{1}     & \ctrl{1}     & \ctrl{1}     & \ctrl{1}     & \ctrl{1}     & \ctrl{1}     & \ctrl{1}     & \gate{R_y}        & \qw \\
& \ctrlo{1}    & \ctrlo{1}    & \ctrl{1}     & \ctrl{1}     & \ctrlo{1}    & \ctrlo{1}    & \ctrl{1}     & \ctrl{1}     & \ctrlo{1}    & \ctrlo{1}    & \ctrl{1}     & \ctrl{1}     & \ctrlboth{-1}     & \qw \\
& \ctrlo{1}    & \ctrl{1}     & \ctrlo{1}    & \ctrl{1}     & \gate{R_y}   & \gate{R_y}   & \gate{R_y}   & \gate{R_y}   & \ctrlo{1}    & \ctrl{1}     & \ctrlo{1}    & \ctrl{1}     & \ctrlboth{-1}     & \qw \\
& \gate{R_y}   & \gate{R_y}   & \gate{R_y}   & \gate{R_y}   & \ctrlo{-1}   & \ctrl{-1}    & \ctrlo{-1}   & \ctrl{-1}    & \gate{R_y}   & \gate{R_y}   & \gate{R_y}   & \gate{R_y}   & \ctrlboth{-1}     & \qw \\
&              & \tilde{A}_{11}^{4} &        &              &              & \tilde{A}_{10}^{3} &        &              &              & \tilde{A}_{9}^{4} &         &              & \tilde{A}_{8}^{1} &     \\
}
\]
\caption{\label{fig:example star}The circuit of $U$, the random walk evolution on the star graph (Fig.~\ref{fig:star}).}
\end{figure*}
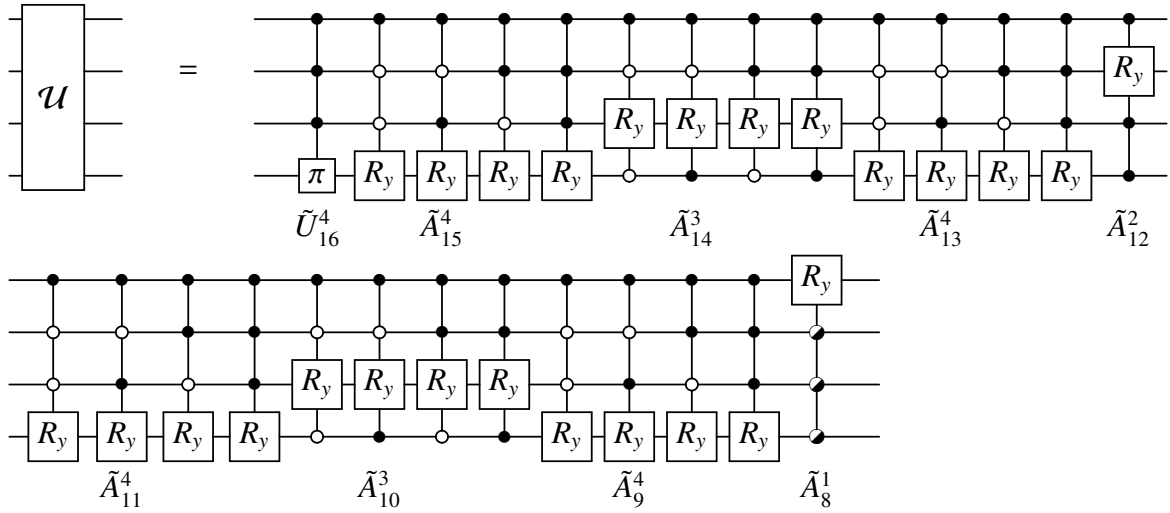
\end{center}

It requires only 34 quantum gates to implement the $16\times 16$ unitary matrix. As a comparison, for a random $16\times 16$ real matrix, it needs around $8\times 15+15=135$ gates, while for a random $16\times 16$ complex matrix, it needs $8\times15+8\times15+16=256$ gates. The last 7 ${{R}_{y}}$ gate combinations (56 ${{R}_{y}}$ subgates) completely disappear. This is another example demonstrating the efficiency of \emph{Qcompiler}.

\subsection{\label{sec:example complicated}Random walk on a complicated graph}

\begin{figure}[!h]
\centering
\includegraphics[scale=0.5]{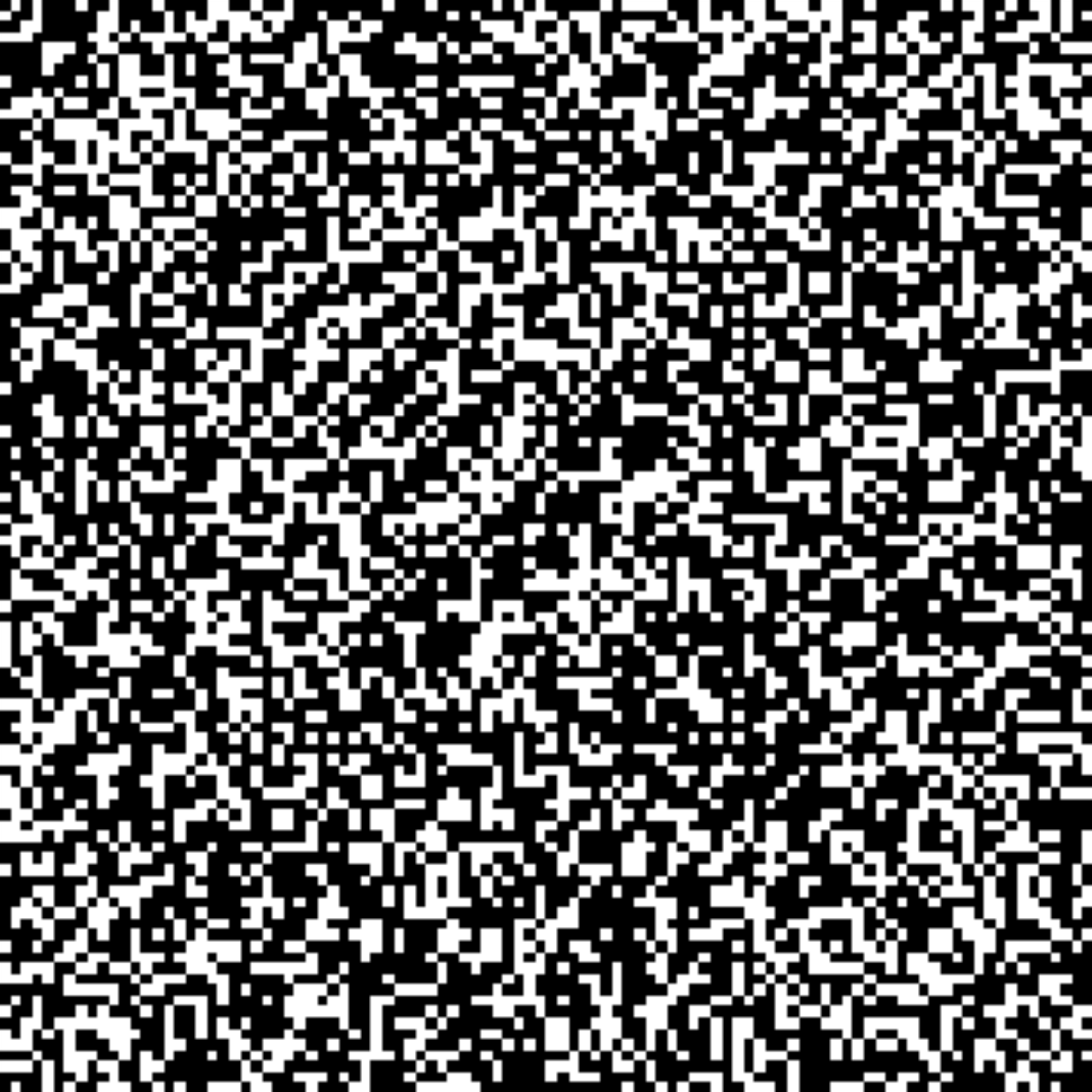}
\caption{\label{fig:adj}The adjacent matrix of a complicated graph.}
\end{figure}

\emph{Qcompiler} also works for very complicated graphs. Fig.~\ref{fig:adj} is an adjacent matrix of a 100-node graph with white dots standing for '1' and black dots for '0'.

The quantum walk operator for this graph is a $4011\times 4011$ unitary matrix $U$.  We first expand the size of the unitary matrix with an identity matrix, i.e.
\begin{equation}
W=\left( \begin{matrix}
   U & {}  \\
   {} & I  \\
\end{matrix} \right)
\end{equation}
where $U$ is the $4011\times 4011$ unitary matrix, $I$ is a $85\times 85$ identity matrix, and $W$ is a ${{2}^{12}}\times {{2}^{12}}$ unitary matrix.  The resulting quantum circuit contains 12 qubits.


\section{\label{sec:conclusion}Conclusions}

We have developed an efficient and versatile package, \emph{Qcompiler}, which maps any unitary matrix $U$ of arbitrary size into a quantum circuit with only one- and two-qubit logic gates by applying cosine-sine decomposition recursively.  For real unitary matrices, \emph{Qcompiler} provides a much simpler quantum circuit with only half number of elementary gates in comparison with earlier work by Mottonen \emph{et al} \cite{Mottonen2004} which deals with general unitary matrices.

The quantum circuits produced by \emph{Qcompiler} also reflects the symmetry of the systems under study.  In particular, we examined the resulting quantum circuits corresponding to quantum walks on graphs with certain degree of symmetry \cite{Douglas2008}.  In this case, many gates in such a circuit turned out to be an identity gate or a simple $\pi$ gate, which can be easily eliminated or combined with other gates to further reduce the complexity of the final quantum circuit. For unitary matrices of certain structure or symmetry, it may be possible to include qutris and qudits to produce simpler and more efficient quantum circuits.  This will be an interesting subject for further study. 


\section*{Acknowledgements}

The authors would like to thank Sergey Koulikov, Blake Segler, and Gary Allwood for their earlier involvement in this project.  Thanks are also due to Kia Manouchehri, Brendan Douglas and Thomas Loke for several valuable discussions.



\bibliographystyle{cpc}
\bibliography{Reference}

\begin{thebibliography}{10}

\bibitem{Deutsch1989}
Deutsch, D.,
\newblock Proceedings of the Royal Society of London Series A - Mathematical
  Physical and Engineering Sciences {\bf 425} (1989) 73.

\bibitem{Shor1997}
Shor, P.~W.,
\newblock SIAM Journal on Computing {\bf 26} (1997) 1484.

\bibitem{Nielsen2000}
Nielsen, M. and Chuang, I.,
\newblock {\em Quantum computation and quantum information},
\newblock Cambridge, Cambridge, 2000.

\bibitem{Loke2011}
Loke, T. and Wang, J.~B.,
\newblock Computer Physics Communications {\bf 182} (2011) 2285Ð2294.

\bibitem{Barenco1995}
Barenco, A. et~al.,
\newblock Physical Review A {\bf 52} (1995) 3457.

\bibitem{Deutsch1995}
Deutsch, D., Barenco, A., and Ekert, A.,
\newblock Proceedings of the Royal Society of London Series a-Mathematical
  Physical and Engineering Sciences {\bf 449} (1995) 669.

\bibitem{Cybenko2001}
Cybenko, G.,
\newblock Computing in Science and Engineering {\bf 3} (2001) 27.

\bibitem{Tucci1999}
Tucci, R.~R.,
\newblock quant-ph/9902062 (1999), quant-ph/0411027 (2004), quant-ph/0407215
  (2004) .

\bibitem{Mottonen2004}
Mottonen, M., Vartiainen, J.~J., Bergholm, V., and Salomaa, M.~M.,
\newblock Physical Review Letters {\bf 93} (2004) 130502.

\bibitem{Bergholm2005}
Bergholm, V., Vartiainen, J.~J., Mottonen, M., and Salomaa, M.~M.,
\newblock Physical Review A {\bf 71} (2005) 052330.

\bibitem{Shende2005}
Shende, V.~V. and Markov, I.~L.,
\newblock Quantum Information \& Computation {\bf 5} (2005) 49.

\bibitem{Khan2006}
Khan, F.~S. and Perkowski, M.,
\newblock Theoretical Computer Science {\bf 367} (2006) 336.

\bibitem{Manouchehri2009}
Manouchehri, K. and Wang, J.~B.,
\newblock Physical Review A {\bf 80} (2009) 060304.

\bibitem{DeVos2009}
De~Vos, A. and Van~Rentergem, Y.,
\newblock Journal of Multiple-Valued Logic and Soft Computing {\bf 15} (2009)
  489.

\bibitem{DeVos2011}
De~Vos, A., Boes, M., and De~Baerdemacker, S.,
\newblock Journal of Multiple-Valued Logic and Soft Computing {\bf 18} (2011)
  67.

\bibitem{Paige1994}
Paige, C.~C. and Wei, M.,
\newblock Linear Algebra and Its Applications {\bf 209} (1994) 303.

\bibitem{Sutton2009}
Sutton, B.,
\newblock Numerical Algorithms {\bf 50} (2009) 33.

\bibitem{Kempe2003}
Kempe, J.,
\newblock Contemporary Physics {\bf 44} (2003) 307.

\bibitem{Douglas2008}
Douglas, B.~L. and Wang, J.~B.,
\newblock J. Phys. A: Math. Theor {\bf 41} (2008) 075303.

\bibitem{Berry2011}
Berry, S.~D. and Wang, J.~B.,
\newblock Physical Review A {\bf 83} (2011) 042317.

\bibitem{Douglas2009}
Douglas, B.~L. and Wang, J.~B.,
\newblock Physical Review A {\bf 79} (2009) 052335.

\bibitem{Shenvi2003}
Shenvi, N., Kempe, J., and Whaley, K.~B.,
\newblock Physical Review A {\bf 67} (2003) 052307.

\end{thebibliography}


\end{document}